\documentclass[11pt,onecolumn,draftcls]{IEEEtran}

\usepackage{amsmath}
\usepackage{graphicx}
\usepackage{amssymb}
\usepackage{setspace}
\usepackage{algorithmic}
\usepackage{algorithm}
\usepackage[tight,footnotesize]{subfigure}

\newtheorem{thm}{Theorem}[section]
\newtheorem{cor}[thm]{Corollary}
\newtheorem{lem}[thm]{Lemma}
\newtheorem{prop}[thm]{Proposition}

\numberwithin{equation}{section}
\newcommand{\ceil}[1]{\left\lceil#1\right\rceil}
\newcommand{\norm}[1]{\left\Vert#1\right\Vert}
\newcommand{\abs}[1]{\left\vert#1\right\vert}
\newcommand\vect[1]{{\bf#1}}
\newcommand\matr[1]{{\bf#1}}
\newcommand\epsilonbf{{\boldsymbol{\epsilon}}}

\newcommand{\argmax}{\operatornamewithlimits{argmax}}

\newcommand{\Real}{\mathbb R}

\DeclareMathOperator{\supp}{supp}
\DeclareMathOperator{\resid}{resid}
\DeclareMathOperator{\proj}{proj}
\DeclareMathOperator{\trace}{trace}
\DeclareMathOperator{\MSE}{MSE}
\DeclareMathOperator{\spanned}{span}

\begin{document}

\title{RIP-Based Near-Oracle Performance Guarantees for Subspace-Pursuit,
CoSaMP, and Iterative Hard-Thresholding}

\author{Raja~Giryes, and Michael~Elad
\thanks{R. Giryes and M. Elad are with the Department of Computer Science,
Technion - IIT 32000, Haifa, ISRAEL}}

\author{Raja Giryes and Michael Elad \\
{\small Department of Computer Science --
Technion -- Israel Institute of Technology} \\
{\small Haifa 32000, Israel} \\
{\small \{raja,elad\}@cs.technion.ac.il}}

\markboth{Preprint}%
{R.~Giryes and M.~Elad}

\maketitle
\begin{abstract}
This paper presents an average case denoising performance analysis
for the Subspace Pursuit (SP), the CoSaMP and the IHT algorithms.
This analysis considers the recovery of a noisy signal, with the
assumptions that (i) it is corrupted by an additive random white
Gaussian noise; and (ii) it has a $K$-sparse representation with
respect to a known dictionary $\matr{D}$. The proposed analysis is
based on the Restricted-Isometry-Property (RIP), establishing a
near-oracle performance guarantee for each of these algorithms. The
results for the three algorithms differ in the bounds' constants and
in the cardinality requirement (the upper bound on $K$ for which the
claim is true).

Similar RIP-based analysis was carried out previously for the
Dantzig Selector (DS) and the Basis Pursuit (BP). Past work also
considered a mutual-coherence-based analysis of the denoising
performance of the DS, BP, the Orthogonal Matching Pursuit (OMP) and
the thresholding algorithms. This work differs from the above as it
addresses a different set of algorithms. Also, despite the fact that
SP, CoSaMP, and IHT are greedy-like methods, the performance
guarantees developed in this work resemble those obtained for the
relaxation-based methods (DS and BP), suggesting that the
performance is independent of the sparse representation entries
contrast and magnitude.
\end{abstract}


\section{Introduction}

\subsection{General -- Pursuit Methods for Denoising}

The area of sparse approximation (and compressed sensing as one
prominent manifestation of its applicability) is an emerging field
that get much attention in the last decade. In one of the most basic
problems posed in this field, we consider a noisy measurement vector
$\vect{y}\in \Real^{m}$ of the form
\begin{equation}
\label{eq:meas_vec}
\vect{y}=\matr{D}\vect{x}+\vect{e},
\end{equation}
where $\vect{x}\in\Real^N$ is the signal's representation with
respect to the dictionary $\matr{D}\in \Real^{m \times N} $ where $N
\ge m$. The vector $\vect{e}\in\Real^m$ is an additive noise, which
is assumed to be an adversial disturbance, or a random vector --
e.g., white Gaussian noise with zero mean and variance $\sigma^2$.
We further assume that the columns of $\matr{D}$ are normalized, and
that the representation vector $\vect{x}$ is $K$-sparse, or nearly
so.\footnote{A more exact definition of nearly sparse vectors will
be given later on} Our goal is to find the $K$-sparse vector
$\vect{x}$ that approximates the measured signal $\vect{y}$. Put
formally, this reads
\begin{equation}
\label{eq:formal_def}
\min_{\vect{x}}~\norm{\vect{y}-\matr{D}\vect{x}}_2~~\text{  subject
to  }\norm{\vect{x}}_0=K,
\end{equation}
where $\norm{\vect{x}}_0$ is the $\ell_0$ pseudo-norm that counts
the number of non-zeros in the vector $\vect{x}$. This problem is
quite hard and problematic
\cite{Donoho06Stable,Donoho06OnTheStability,Tropp06JustRelax,Bruckstein09From}.
A straight forward search for the solution of (\ref{eq:formal_def})
is an NP hard problem as it requires a  combinatorial search over
the support of $\vect{x}$ \cite{NP-Hard}. For this reason,
approximation algorithms were proposed -- these are often referred
to as pursuit algorithms.

One popular pursuit approach is based on $\ell_1$ relaxation and
known as the Basis Pursuit (BP) \cite{Chen98overcomplete} or the
Lasso \cite{Tibshirani96Regression}. The BP aims at minimizing the
relaxed objective
\begin{eqnarray}
\label{eq:p_1} (P1):& \min_{\vect{x}} \norm{\vect{x}}_1 \text{ subject to } \norm{\vect{y} - \matr{D}\vect{x}}_2^2 \le
\epsilon_\vect{BP}^2,
\end{eqnarray}
where $\epsilon_\vect{BP}$ is a constant related to the noise power.
This minimizing problem has an equivalent form:
\begin{eqnarray}
\label{eq:bp} (BP):& \min_{\vect{x}} & \frac{1}{2}\norm{\vect{y} -
\matr{D}\vect{x}}_2^2 + \gamma_{BP} \norm{\vect{x}}_1,
\end{eqnarray}
where $\gamma_{BP}$ is a constant related to $\epsilon_\vect{BP}$.
Another $\ell_1$-based relaxed algorithm is the Dantzig Selector
(DS), as proposed in \cite{Candes07Dantzig}. The DS aims at
minimizing
\begin{eqnarray}
\label{eq:ds} (DS): \min_{\vect{x}} \norm{\vect{x}}_1 \text{
subject to } \norm{\matr{D}^*(\vect{y} - \matr{D}\vect{x})}_\infty
\le \epsilon_\vect{DS},
\end{eqnarray}
where $\epsilon_\vect{DS}$ is a constant related to the noise power.

A different pursuit approach towards the approximation of the
solution of (\ref{eq:formal_def}) is the greedy strategy
\cite{Chen89Orthogonal,MallatZhang93,Davis97Adaptive-greedy},
leading to algorithms such as the Matching Pursuit (MP) and the
Orthogonal Matching Pursuit (OMP). These algorithms build the
solution $\vect{x}$ one non-zero entry at a time, while greedily
aiming to reduce the residual error $\norm{\vect{y} -
\matr{D}\vect{x}}_2^2$.

The last family of pursuit methods we mention here are greedy-like
algorithms that differ from MP and OMP in two important ways: (i)
Rather than accumulating the desired solution one element at a time,
a group of non-zeros is identified together; and (ii) As opposed to
the MP and OMP, these algorithms enable removal of elements from the
detected support. Algorithms belonging to this group are the
Regularized OMP (ROMP) \cite{Needell10Signal}, the Compressive
Sampling Matching Pursuit (CoSaMP) \cite{Needell09CoSaMP}, the
Subspace-Pursuit (SP) \cite{Dai09Subspace}, and the Iterative Hard
Thresholding (IHT) \cite{Blumensath09Iterative}. This paper focuses
on this specific family of methods, as it poses an interesting
compromise between the simplicity of the greedy methods and the
strong abilities of the relaxed algorithms.

\subsection{Performance Analysis -- Basic Tools}

Recall that we aim at recovering the (deterministic!) sparse
representation vector $\vect{x}$. We measure the quality of the
approximate solution $\hat{\vect{x}}$ by the Mean-Squared-Error
(MSE)
\begin{eqnarray}
\label{eq:objective} \MSE(\hat{\vect{x}}) = E\norm{\vect{x}
-\hat{\vect{x}}}_2^2,
\end{eqnarray}
where the expectation is taken over the distribution of the noise.
Therefore, our goal is to get as small as possible error. The
question is, how small can this noise be? In order to answer this
question, we first define two features that characterize the
dictionary $\matr{D}$ -- the mutual coherence and the Restricted
Isometry Property (RIP). Both are used extensively in formulating
the performance guarantees of the sort developed in this paper.

The mutual-coherence $\mu$
\cite{Donoho01Uncertainty,Elad02generalized,Donoho03Optimal} of a
matrix $\matr{D}$ is the largest absolute normalized inner product
between different columns from $\matr{D}$. The larger it is, the
more problematic the dictionary is, because in such a case we get
that columns in $\matr{D}$ are too much alike.

Turning to the RIP, it is said that $\matr{D}$ satisfies the $K$-RIP
condition with parameter $\delta_K$ if it is the smallest value that
satisfies
\begin{equation}
(1-\delta_K) \norm{\vect{x}}_2^2 \le \norm{\matr{D}\vect{x}}_2^2 \le
(1+\delta_K) \norm{\vect{x}}_2^2
\end{equation}
for any $K$-sparse vector $\vect{x}$ \cite{Candes06Near,Candes05Decoding}.

These two measures are related by $\delta_K \le (K-1)\mu$
\cite{BenHaim09Coherence}. The RIP is a stronger descriptor of
$\matr{D}$ as it characterizes groups of $K$ columns from
$\matr{D}$, whereas the mutual coherence ``sees'' only pairs. On the
other hand, computing $\mu$ is easy, while the evaluation of
$\delta_K$ is prohibitive in most cases. An exception to this are
random matrices $\matr{D}$ for which the RIP constant is known (with
high probability). For example, if the entries of $\sqrt{m}\matr{D}$
are drawn from a white Gaussian distribution\footnote{The
multiplication by $\sqrt{m}$ comes to normalize the columns of the
effective dictionary $\matr{D}$.} and $m \ge
CK\log(N/K)/\epsilon^2$,
then with a very high probability $\delta_K \le \epsilon$
\cite{Candes06Near,Rudelson06Sparse}.

We return now to the question we posed above: how small can the
error $\MSE(\hat{\vect{x}})$ be? Consider an oracle estimator that
knows the support of $\vect{x}$, i.e. the locations of the $K$
non-zeros in this vector. The oracle estimator obtained as a direct
solution of the problem posed in (\ref{eq:formal_def}) is easily
given by
\begin{eqnarray}
\label{eq:oracle} \hat{\vect{x}}_{oracle} = \matr{D}_T^\dag
\vect{y},
\end{eqnarray}
where $T$ is the support of $\vect{x}$ and $\matr{D}_T$ is a
sub-matrix of $\matr{D}$ that contains only the columns involved in
the support $T$. Its MSE is given by \cite{Candes07Dantzig}
\begin{eqnarray}
\label{eq:oracle_perf}\MSE(\hat{\vect{x}}_{oracle}) =
E\norm{\vect{x} - \hat{\vect{x}}_{oracle}}_2^2 =
E\norm{\matr{D}_T^\dag \vect{e}}_2^2.
\end{eqnarray}
In the case of a random noise, as described above, this error
becomes
\begin{eqnarray}
\label{eq:oracle_perf2} \MSE  (\hat{\vect{x}}_{oracle}) &=&
E\norm{\matr{D}_T^\dag \vect{e}}_2^2  \\
\nonumber
& =& \trace\left\{(\matr{D}^*_T\matr{D}_T)^{-1}\right\} \sigma^2 \\
\nonumber
&\le& \frac{K}{1-\delta_K} \sigma^2.
\end{eqnarray}
This is the smallest possible error, and it is proportional to the
number of non-zeros $K$ multiplied by $\sigma^2$. It is natural to
ask how close do we get to this best error by practical pursuit
methods that do not assume the knowledge of the support. This brings
us to the next sub-section.

\subsection{Performance Analysis -- Known Results}

There are various attempts to bound the MSE of pursuit algorithms.
Early works considered the adversary case, where the noise can admit
any form as long as its norm is bounded \cite{Tropp06Just,
Candes06Stable,Donoho06OnTheStability,Donoho06Stable}. These works
gave bounds on the reconstruction error in the form of a constant
factor ($Const> 1$) multiplying the noise power,
\begin{eqnarray}
\label{eq:perf_Advers} \norm{\vect{x} - \hat{\vect{x}}}^2_2 \le
Const \cdot \norm{\vect{e}}_2^2.
\end{eqnarray}
Notice that the cardinality of the representation plays no role in
this bound, and all the noise energy is manifested in the final
error.

One such example is the work by Cand\`{e}s and Tao, reported in
\cite{Candes05Decoding}, which analyzed the BP error. This work have
shown that if the dictionary $\matr{D}$ satisfies $\delta_{K}
+\delta_{2K} + \delta_{3K} < 1$ then the BP MSE is bounded by a
constant times the energy of the noise, as shown above. The
condition on the RIP was improved to $\delta_{2K} < \sqrt{2} -1 $ in
\cite{Candes08Comptes}. Similar tighter bounds are $\delta_{1.625K}
< \sqrt{2} - 1$ and $\delta_{3K} < 4 - 2\sqrt{3} $
\cite{Cai01Shifting}, or $\delta_{K} <0.307$ \cite{Cai01New}. The
advantage of using the RIP in the way described above is that it
gives a uniform guarantee: it is related only to the dictionary and
sparsity level.

Next in line to be analyzed are the greedy methods (MP, OMP, Thr)
\cite{Tropp06Just,Donoho06Stable}. Unlike the BP, these algorithms
where shown to be more sensitive, incapable of providing a uniform
guarantee for the reconstruction. Rather, beyond the dependence on
the properties of $\matr{D}$ and the sparsity level, the guarantees
obtained depend also on the ratio between the noise power and the
absolute values of the signal representation entries.

Interestingly, the greedy-like approach, as practiced in the ROMP,
the CoSaMP, the SP, and the IHT algorithms, was found to be closer
is spirit to the BP, all leading to uniform guarantees on the
bounded MSE. The ROMP was the first of these algorithms to be
analyzed \cite{Needell10Signal}, leading to the more strict
requirement $\delta_{2K}<0.03/\sqrt{\log K}$. The CoSaMP
\cite{Needell09CoSaMP} and the SP \cite{Dai09Subspace} that came
later have similar RIP conditions without the $\log K$ factor, where
the SP result is slightly better.
The IHT algorithm was also shown to have a uniform guarantee for
bounded error of the same flavor as shown above
\cite{Blumensath09Iterative}.

All the results mentioned above deal with an adversial noise, and
therefore give bounds that are related only to the noise power with
a coefficient that is larger than $1$, implying that no effective
denoising is to be expected. This is natural since we consider the
worst case results, where the noise can be concentrated in the
places of the non-zero elements of the sparse vector. To obtain
better results, one must change the perspective and consider a
random noise drawn from a certain distribution.

The first to realize this and exploit this alternative point of view
were Candes and Tao in the work reported in \cite{Candes07Dantzig}
that analyzed the DS algorithm. As mentioned above, the noise was
assumed to be random zero-mean white Gaussian noise with a known
variance $\sigma^2$. For the choice $\epsilon_{DS} = \sqrt{2(1+a)
\log{N}}\cdot\sigma$, and requiring $\delta_{2K} +\delta_{3K} < 1$,
the minimizer of (\ref{eq:ds}), $\hat{\vect{x}}_{DS}$, was shown to
obey
\begin{eqnarray}
\label{eq:ds_perf} \norm{\vect{x} - \hat{\vect{x}}_{DS}}^2_2 \le
C_{DS}^2 \cdot (2 (1+a) \log N) \cdot K \sigma^2,
\end{eqnarray}
with probability exceeding $1- (\sqrt{\pi (1+a)\log N}\cdot N^a)^{-1}$,
where $C_{DS} = 4/(1-2\delta_{3K})$.\footnote{In
\cite{Candes07Dantzig} a slightly different constant was presented.}
Up to a constant and a $\log N$ factor, this bound is the same as
the oracle's one in (\ref{eq:oracle_perf}). The $\log N$ factor in
(\ref{eq:ds_perf}) in unavoidable, as proven in
\cite{Candes06Modern}, and therefore this bound is optimal up to a
constant factor.

A similar result was presented in \cite{Bickel09Simultaneous} for
the BP, showing that the solution of (\ref{eq:bp}) for the choice
$\gamma_{BP} = \sqrt{8\sigma^2(1+a)\log N}$, and requiring
$\delta_{2K} + 3\delta_{3K} < 1$, satisfies
\begin{eqnarray}
\label{eq:bp_perf} \norm{\vect{x} - \hat{\vect{x}}_{DS}}^2_2 \le
C_{BP}^2 \cdot (2 (1+a) \log N) \cdot K \sigma^2
\end{eqnarray}
with probability exceeding $1- (N^a)^{-1}$. This result is weaker
than the one obtained for the DS in three ways: (i) It gives a
smaller probability of success; (ii) The constant $C_{BP}$ is
larger, as shown in \cite{BenHaim09Coherence} ($C_{BP} \ge
32/\kappa^4$, where $\kappa < 1$ is defined in
\cite{Bickel09Simultaneous}); and (iii) The condition on the RIP is
stronger.

Mutual-Coherence based results for the DS and BP were derived in
\cite{Cai09Stable,BenHaim09Coherence}. In \cite{BenHaim09Coherence}
results were developed also for greedy algorithms -- the OMP and the
thresholding. These results rely on the contrast and magnitude of
the entries of $\vect{x}$. Denoting by $\hat{\vect{x}}_{greedy}$ the
reconstruction result of the thresholding and the OMP, we have
\begin{eqnarray}
\label{eq:greedy_perf} \norm{\vect{x} - \hat{\vect{x}}_{greedy}}^2_2
\le C_{greedy}^2 \cdot (2 (1+a) \log N) \cdot K \sigma^2,
\end{eqnarray}
where $C_{greedy} \le 2$ and with probability  exceeds $1-
(\sqrt{\pi (1+a) \log N}\cdot N^a)^{-1}$. This result is true for
the OMP and thresholding under the condition
\begin{eqnarray}
\label{eq:OMP_and_THR_cond} \frac{\abs{\vect{x}_{min}} - 2\sigma
\sqrt{2(1+a)\log N }}{(2K-1)\mu}  \ge \left\{
\begin{array}{cc}\abs{ \vect{x}_{min}} & \text{OMP} \\
|\vect{x}_{max}| & \text{THR} \end{array}\right.,
\end{eqnarray}
where $|\vect{x}_{min}|$ and $|\vect{x}_{max}|$ are the minimal and
maximal non-zero absolute entries in $\vect{x}$.

\subsection{This Paper Contribution}

We have seen that greedy algorithms' success is dependent on the
magnitude of the entries of $\vect{x}$ and the noise power, which is
not the case for the DS and BP. It seems that there is a need for
pursuit algorithms that, on one hand, will enjoy the simplicity and
ease of implementation as in the greedy methods, while being
guaranteed to perform as well as the BP and DS. Could the
greedy-like methods (ROMP, CoSaMP, SP, IHT) serve this purpose? The
answer was shown to be positive for the adversial noise assumption,
but these results are too weak, as they do not show the true
denoising effect that such algorithm may lead to. In this work we
show that the answer remains positive for the random noise
assumption.

More specifically, in this paper we present RIP-based near-oracle
performance guarantees for the SP, CoSaMP and IHT algorithms (in
this order). We show that these algorithms get uniform guarantees,
just as for the relaxation based methods (the DS and BP). We present
the analysis that leads  to these results and we provide explicit
values for the constants in the obtained bounds.

The organization of this paper is as follows: In Section
\ref{sec:notation} we introduce the notation and propositions used
for our analysis. In Section~\ref{sec:near_orac_perf} we develop RIP-based
bounds for the SP, CoSaMP and the IHT algorithms for the adversial case.
Then we show how we can derive from these a new set of guarantees for near
oracle performance that consider the noise as random.
We develop fully the steps for the SP, and
outline the steps needed to get the results for the CoSaMP and IHT.
In Section \ref{sec:experiments} we present some experiments that
show the performance of the three methods, and a comparison between
the theoretical bounds and the empirical performance. In Section
\ref{sec:non_exact_sparse} we consider the nearly-sparse case, extending
all the above results. Section \ref{sec:conc} concludes our work.


\section{Notation and Preliminaries}
\label{sec:notation} The following notations are used in this paper:
\begin{itemize}
\item $\supp(\vect{x})$ is the support of $\vect{x}$ (a set with the locations of
the non-zero elements of $\vect{x}$).
\item $\abs{\supp(\vect{x})}$ is the size of the set $\supp(\vect{x})$.
\item  $\supp(\vect{x},K)$ is the support of the largest $K$
magnitude elements in $\vect{x}$.
\item $\matr{D}_T$ is a matrix composed of the columns of the
matrix $\matr{D}$ of the set $T$.
\item In a similar way, $\vect{x}_T$ is a vector composed of the entries
of the vector $\vect{x}$ over the set $T$.
\item $T^C$ symbolizes the complementary set of $T$.
\item $T - \tilde{T}$ is the set of all the elements contained
in $T$ but not in $\tilde{T}$.
\item We will denote by $T$ the set of the non-zero places of
the original signal $\vect{x}$; As such, $\abs{T} \le K$ when
$\vect{x}$ is $K$-sparse.
\item $\vect{x}_K$ is the vector with the $K$ dominant elements
of $\vect{x}$.
\item The projection of a vector $\vect{y}$ to the subspace
spanned by the columns of the matrix $\matr{A}$ (assumed to have
more rows than columns) is denoted by $\proj(\vect{y},\matr{A}) =
\matr{A}\matr{A}^\dag\vect{y}$. The residual is
$\resid(\vect{y},\matr{A}) = \vect{y} -
\matr{A}\matr{A}^\dag\vect{y}$.
\item $T_e$ is the subset of columns of size $K$ in $\matr{D}$ that
gives the maximum correlation with the noise vector $\vect{e}$,
namely,
\begin{equation}
\label{eq:T_e_def} T_e = \argmax_{T\left|~|T|=K \right.}
\norm{\matr{D}_{T}^*\vect{e}}_2
\end{equation}
\item  $T_{\vect{e},p}$ is a generalization of $T_{\vect{e}}$
where $T$ in (\ref{eq:T_e_def}) is of size $pK$, $p \in \mathbb{N}$.
It is clear that $\norm{\matr{D}_{T_{\vect{e},p}}^*\vect{e}}_2 \le
p\norm{\matr{D}_{T_{\vect{e}}}^*\vect{e}}_2$.
\end{itemize}

\noindent The proofs in this paper use several propositions from
\cite{Needell09CoSaMP, Dai09Subspace}. We bring these in this
Section, so as to keep the discussion complete.

\begin{prop}\label{prop1}[Proposition 3.1 in \cite{Needell09CoSaMP}]
Suppose $\matr{D}$ has a restricted isometry constant $\delta_{K}$.
Let $T$ be a set of $K$ indices or fewer. Then
\begin{eqnarray*}
\norm{\matr{D}_T^*\vect{y}}_2 \le \sqrt{1+\delta_{K}}\norm{\vect{y}}_2 \\
\norm{\matr{D}_T^\dag\vect{y}}_2 \le \frac{1}{\sqrt{1-\delta_{K}}}\norm{\vect{y}}_2 \\
\norm{\matr{D}_T^*\matr{D}_T\vect{x}}_2 \lesseqqgtr (1\pm\delta_{K})\norm{\vect{x}}_2 \\
\norm{(\matr{D}_T^*\matr{D}_T)^{-1}\vect{x}}_2 \lesseqqgtr
\frac{1}{1\pm\delta_{K}}\norm{\vect{x}}_2
\end{eqnarray*}
where the last two statements contain upper and lower bounds,
depending on the sign chosen.
\end{prop}

\begin{prop}\label{prop2}[Lemma 1 in \cite{Dai09Subspace}] Consequences of the RIP:
\begin{enumerate}
\item (Monotonicity of $\delta_K$) For any two integers $K\le K'$,
$\delta_{K} \le \delta_{K'}$.
\item  (Near-orthogonality of columns) Let $I,J \subset \{1,...,N\}$ be
two disjoint sets ($I \cap J = \emptyset$). Suppose that
$\delta_{\abs{I}+\abs{J}} < 1$. For arbitrary vectors $\vect{a} \in
\Real^{\abs{I}}$ and $\vect{b}\in \Real^{\abs{J}}$,
\begin{eqnarray}
\nonumber \abs{\langle \matr{D}_I \vect{a}, \matr{D}_J\vect{b}
\rangle} \le \delta_{\abs{I}+\abs{J}} \norm{\vect{a}}_2
\norm{\vect{b}}_2
\end{eqnarray}
and
\begin{eqnarray}
\nonumber
\norm{\matr{D}_I^*\matr{D}_J\vect{b}}_2 \le \delta_{\abs{I}+\abs{J}}
\norm{\vect{b}}_2.
\end{eqnarray}
\end{enumerate}
\end{prop}

\begin{prop}\label{prop3}[Lemma 2 in \cite{Dai09Subspace}] Projection and
Residue:
\begin{enumerate}
\item (Orthogonality of the residue) For an arbitrary
vector $\vect{y} \in \Real^m$ and a sub-matrix $\matr{D}_I \in
\Real^{m \times K}$ of full column-rank, let $\vect{y}_r =
\resid(\vect{y,\matr{D}_I})$. Then $\matr{D}^*_I\vect{y}_r = 0$.
\item (Approximation of the projection residue)
Consider a matrix $\matr{D} \in \Real^{m \times N}$. Let $I,J
\subset \{1,...,N\}$ be two disjoint sets, $I\cap J = \emptyset$,
and suppose that $\delta_{\abs{I}+\abs{J}} < 1$. Let $\vect{y} \in
\spanned(\matr{D}_I)$, $\vect{y}_p = \proj(\vect{y},\matr{D}_J)$ and
$\vect{y}_r = \resid(\vect{y},\matr{D}_J)$. Then
\begin{eqnarray*}
\norm{\vect{y}_p}_2 \le \frac{\delta_{\abs{I}+\abs{J}}}{1-\delta_{\max(\abs{I},\abs{J})}}\norm{\vect{y}}_2
\end{eqnarray*}
and
\begin{eqnarray*}
\left( 1 -
\frac{\delta_{\abs{I}+\abs{J}}}{1-\delta_{\max(\abs{I},\abs{J})}}
\right)\norm{\vect{y}}_2 \le \norm{\vect{y}_r}_2 \le
\norm{\vect{y}}_2.
\end{eqnarray*}
\end{enumerate}
\end{prop}

\begin{prop}\label{prop4}[Corollary 3.3 in \cite{Needell09CoSaMP}]
Suppose that $\matr{D}$ has an RIP constant $\delta_{\tilde{K}}$.
Let $T_1$ be an arbitrary set of indices, and let $\vect{x}$ be a
vector. Provided that $\tilde{K} \ge \abs{T_1\cup \supp(\vect{x})}$,
we obtain that
\begin{eqnarray}
\norm{\matr{D}^*_{T_1} \matr{D}_{T^C_1}\vect{x}_{T^C_1}}_2 \le
\delta_{\tilde{K}} \norm{\vect{x}_{T^C_1}}_2.
\end{eqnarray}
\end{prop}


\section{Near oracle performance of the algorithms}
\label{sec:near_orac_perf}

Our goal in this section is to find error bounds for the SP, CoSaMP
and IHT reconstructions given the measurement from
(\ref{eq:meas_vec}). We will first find bounds for the case where
$\vect{e}$ is an adversial noise using the same techniques used in
\cite{Dai09Subspace,Needell09CoSaMP}. In these works and in
\cite{Blumensath09Iterative}, the reconstruction error was bounded
by a constant times the noise power in the same form as in
(\ref{eq:perf_Advers}). In this work, we will derive a bound that is
a constant times $\norm{\matr{D}_{T_\vect{e}}^*\vect{e}}_2$ (where
$T_\vect{e}$ is as defined in the previous section). Armed with this
bound, we will change perspective and look at the case where
$\vect{e}$ is a white Gaussian noise, and derive a near-oracle
performance result of the same form as in (\ref{eq:ds_perf}), using
the same tools used in \cite{Candes07Dantzig}.


\subsection{Near oracle performance of the SP algorithm}

\label{sec:SP_perf} We begin with the SP pursuit method, as
described in Algorithm~\ref{alg:SP}. SP holds a temporal solution
with $K$ non-zero entries, and in each iteration it adds an
additional set of $K$ candidate non-zeros that are most correlated
with the residual, and prunes this list back to $K$ elements by
choosing the dominant ones. We use a constant number of iterations
as a stopping criterion but different stopping criteria can be
sought, as presented in \cite{Dai09Subspace}.

\begin{algorithm}
\caption{Subspace Pursuit Algorithm [Algorithm 1 in
\cite{Dai09Subspace}]} \label{alg:SP}
\begin{algorithmic}

\REQUIRE $K, \matr{D}, \vect{y}$ where $\vect{y} = \matr{D}\vect{x}
+ \vect{e}$, $K$ is the cardinality of $\vect{x}$ and $\vect{e}$ is
the additive noise.

\ENSURE $\hat{\vect{x}}_{SP}$: $K$-sparse approximation of
$\vect{x}$

\STATE Initialize the support: $T^0 =\emptyset$.

\STATE Initialize the residual: $\vect{y}_r^0 = \vect{y}$.

\WHILE{halting criterion is not satisfied}

\STATE Find new support elements: $T_\Delta =
\supp(\matr{D}^*\vect{y}^{\ell - 1}_r,K)$.

\STATE Update the support: $\tilde{T}^\ell = T^{\ell -1} \cup
T_\Delta$.

\STATE Compute the representation: $\vect{x}_p =
\matr{D}^{\dag}_{\tilde{T}^\ell}\vect{y}$.

\STATE Prune small entries in the representation: $T^\ell =
\supp(\vect{x}_p,K)$.

\STATE Update the residual: $\vect{y}_r^\ell =
\resid(\vect{y},\matr{D}_{T^\ell})$.

\ENDWHILE \STATE Form the final solution:
$\hat{\vect{x}}_{SP,(T^\ell)^C} = 0$ and
$\hat{\vect{x}}_{SP,T^\ell} = \matr{D}^{\dag}_{T^\ell}\vect{y}$.
\end{algorithmic}
\end{algorithm}

\begin{thm} \label{thm:SP-1} The SP solution at the $\ell$-th iteration
satisfies the recurrence inequality
\begin{eqnarray}
\label{eq:SP_x_diff_bound} \norm{\vect{x}_{T-T^\ell}}_2 &\le&
\frac{2\delta_{3K}(1+\delta_{3K})}{(1-\delta_{3K})^3}
\norm{\vect{x}_{T-T^{\ell-1}}}_2 \\
\nonumber &&+
\frac{6 - 6\delta_{3K} + 4\delta_{3K}^2}
{(1-\delta_{3K})^3}
\norm{\matr{D}_{T_\vect{e}}^*\vect{e}}_2.
\end{eqnarray}
For $\delta_{3K} \le 0.139$ this leads to
\begin{equation}
\label{eq:SP_x_diff_bound_consts} \norm{\vect{x}_{T-T^\ell}}_2 \le
0.5\norm{\vect{x}_{T-T^{\ell-1}}}_2+8.22\norm{\matr{D}_{T_\vect{e}}^*\vect{e}}_2.
\end{equation}
\end{thm}

\begin{figure}[!t]
\begin{center}
\includegraphics[width=0.45\textwidth]{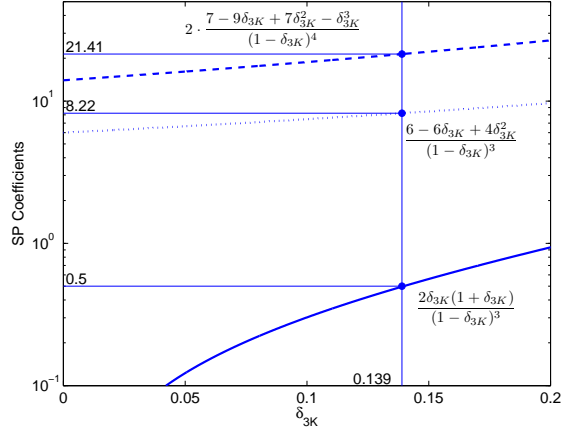}
\end{center}
\caption{The coefficients in (\ref{eq:SP_x_diff_bound})
and (\ref{eq:SP_error_bound_consts}) as functions of $\delta_{3K}$.}
\label{fig:SP-Coefficients}
\end{figure}

{\em Proof:} The proof of the inequality in
(\ref{eq:SP_x_diff_bound}) is given in Appendix
\ref{sec:SP_x_diff_bound_proof}.  Note that the recursive formula
given (\ref{eq:SP_x_diff_bound}) has two coefficients, both
functions of $\delta_{3K}$. Fig.~\ref{fig:SP-Coefficients} shows
these coefficients as a function of $\delta_{3K}$. As can be seen,
under the condition $\delta_{3K} \le 0.139$, it holds that the
coefficient multiplying $\norm{\vect{x}_{T-T^{\ell-1}}}_2$ is lesser
or equal to $0.5$, while the coefficient multiplying
$\norm{\matr{D}_{T_\vect{e}}^*\vect{e}}_2$ is lesser or equal to
$8.22$, which completes our proof. \hfill $\Box$
\bigskip

\begin{cor}
\label{cor:SP_bound_cor} Under the condition $\delta_{3K} \le
0.139$, the SP algorithm satisfies
\begin{equation}
\label{eq:SP_bound_consts} \norm{\vect{x}_{T-T^\ell}}_2 \le
2^{-\ell}\norm{\vect{x}}_2+2\cdot
8.22\norm{\matr{D}_{T_\vect{e}}^*\vect{e}}_2.
\end{equation}
In addition, After at most
\begin{equation}\label{eq:Num_of_Iter_SP}
\ell^* = \ceil{\log_2 \left(
\frac{\norm{\vect{x}}_2}{\norm{\matr{D}_{T_\vect{e}}^*\vect{e}}_2}
\right)}
\end{equation}
iterations, the solution $\hat{\vect{x}}_{SP}$ leads to an accuracy
\begin{equation}
\label{eq:SP_error_bound_consts} \norm{\vect{x} -
\hat{\vect{x}}_{SP}}_2 \le
C_{SP}\norm{\matr{D}_{T_\vect{e}}^*\vect{e}}_2,
\end{equation}
where
\begin{equation}
C_{SP} =  2\cdot \frac{7 -9\delta_{3K}+7\delta_{3K}^2 -
\delta_{3K}^3 }{(1-\delta_{3K})^4} \le 21.41
\end{equation}
\end{cor}

{\em Proof:} Starting with (\ref{eq:SP_x_diff_bound_consts}), and
applying it recursively we obtain
\begin{eqnarray}
&&\norm{\vect{x}_{T-T^\ell}}_2  \le
0.5\norm{\vect{x}_{T-T^{\ell-1}}}_2 +8.22\norm{\matr{D}_{T_\vect{e}}^*\vect{e}}_2
\\ \nonumber  && ~~~~~\le
0.5^2\norm{\vect{x}_{T-T^{\ell-2}}}_2+8.22\cdot(0.5+1)
\norm{\matr{D}_{T_\vect{e}}^*\vect{e}}_2
\\ \nonumber &&  ~~~~~\le  \ldots  \\ \nonumber  && ~~~~~\le
0.5^k\norm{\vect{x}_{T-T^{\ell-k}}}_2+ 8.22\cdot
\left(\sum_{j=0}^{k-1}
0.5^j\right)\norm{\matr{D}_{T_\vect{e}}^*\vect{e}}_2.
\end{eqnarray}
Setting $k=\ell$ leads easily to (\ref{eq:SP_bound_consts}), since
$\norm{\vect{x}_{T-T^{0}}}_2 = \norm{\vect{x}_{T}}_2 =
\norm{\vect{x}}_2$.

Plugging the number of iterations $\ell^*$ as in
 (\ref{eq:Num_of_Iter_SP}) to (\ref{eq:SP_bound_consts})
yields\footnote{Note that we have replaced the constant $8.22$ with
the equivalent expression that depends on $\delta_{3K}$ -- see
 (\ref{eq:SP_x_diff_bound}).}
\begin{eqnarray}\label{eq:temp_bound1}
&& \norm{\vect{x}_{T-T^{\ell^*}}}_2 \\ \nonumber && ~~~~~\le
2^{-\ell^*}\norm{\vect{x}}_2+2\cdot
\frac{6 - 6\delta_{3K} + 4\delta_{3K}^2}{(1-\delta_{3K})^3}
\norm{\matr{D}_{T_\vect{e}}^*\vect{e}}_2
\\ \nonumber  && ~~~~~\le  \left(1+ 2 \cdot \frac{6 - 6\delta_{3K} + 4\delta_{3K}^2}
{(1-\delta_{3K})^3}\right)
\norm{\matr{D}_{T_\vect{e}}^*\vect{e}}_2.
\end{eqnarray}
We define $\hat{T} \triangleq T^{\ell^*}$ and bound the
reconstruction error $\norm{\vect{x} - \hat{\vect{x}}_{SP}}_2$.
First, notice that $\norm{\vect{x}} =
\norm{\vect{x}_{\hat{T}}}+\norm{\vect{x}_{T-\hat{T}}}$, simply
because the true support $T$ can be divided into\footnote{The vector
$\vect{x}_{\hat{T}}$ is of length $|\hat{T}|=K $ and it contains
zeros in locations that are outside $T$.} $\hat T$ and the
complementary part, $T-\hat{T}$. Using the facts that
$\hat{\vect{x}}_{SP} =\matr{D}^\dag_{\hat{T}}\vect{y}$, $\vect{y}=
\matr{D}_T\vect{x}_T+\vect{e}$, and the triangle inequality, we get
\begin{eqnarray}
&&\norm{\vect{x} - \hat{\vect{x}}_{SP}}_2\\ \nonumber && ~~~~~\le
   \norm{\vect{x}_{\hat{T}} - \matr{D}^\dag_{\hat{T}}\vect{y}}_2  +
   \norm{\vect{x}_{T-\hat{T}}}_2\\
\nonumber
&& ~~~~~=  \norm{\vect{x}_{\hat{T}} - \matr{D}^\dag_{\hat{T}}(\matr{D}_T\vect{x}_T+
     \vect{e})}_2  + \norm{\vect{x}_{T-\hat{T}}}_2 \\
\nonumber
&& ~~~~~\le  \norm{\vect{x}_{\hat{T}} - \matr{D}^\dag_{\hat{T}}\matr{D}_T\vect{x}_T}_2
   + \norm{ \matr{D}^\dag_{\hat{T}}\vect{e}}_2+
   \norm{\vect{x}_{T-\hat{T}}}_2.
\end{eqnarray}
We proceed by breaking the term $\matr{D}_T\vect{x}_T$ into the sum
$\matr{D}_{T\cap \hat{T}}\vect{x}_{T\cap \hat{T}} + \matr{D}_{T -
\hat{T}}\vect{x}_{T - \hat{T}}$, and obtain
\begin{IEEEeqnarray}{rCl}
 \norm{\vect{x} - \hat{\vect{x}}_{SP}}_2 & \le& \norm{\vect{x}_{\hat{T}}
- \matr{D}^\dag_{\hat{T}}\matr{D}_{T\cap
\hat{T}}\vect{x}_{T\cap \hat{T}}}_2
\\ \nonumber
&& + \norm{\matr{D}^\dag_{\hat{T}}\matr{D}_{T-\hat{T}}\vect{x}_{T-\hat{T}}}_2
  + \norm{
(\matr{D}^*_{\hat{T}}\matr{D}_{\hat{T}})^{-1}\matr{D}^*_{\hat{T}}\vect{e}}_2+
\norm{\vect{x}_{T-\hat{T}}}_2.
\end{IEEEeqnarray}
The first term in the above inequality vanishes, since
$\matr{D}_{T\cap \hat{T}}\vect{x}_{T\cap \hat{T}} =
\matr{D}_{\hat{T}}\vect{x}_{\hat{T}}$ (recall that
$\vect{x}_{\hat{T}}$ outside the support $T$ has zero entries that
do not contribute to the multiplication). Thus, we get that
$\vect{x}_{\hat{T}} - \matr{D}^\dag_{\hat{T}}\matr{D}_{T\cap
\hat{T}}\vect{x}_{T\cap \hat{T}} = \vect{x}_{\hat{T}} -
\matr{D}^\dag_{\hat{T}}\matr{D}_{\hat{T}}\vect{x}_{\hat{T}}=0$. The
second term can be bounded using Propositions \ref{prop1} and \ref{prop2},
\begin{eqnarray}
\nonumber
&& \norm{\matr{D}^\dag_{\hat{T}}\matr{D}_{T-\hat{T}}\vect{x}_{T-\hat{T}}}_2
=    \norm {(\matr{D}^*_{\hat{T}}\matr{D}_{\hat{T}})^{-1}
\matr{D}^*_{\hat{T}}\matr{D}_{T-\hat{T}}\vect{x}_{T-\hat{T}}}_2 \\
\nonumber && ~~~~~ \le  \frac{1}{1-\delta_K} \norm{
\matr{D}^*_{\hat{T}}\matr{D}_{T-\hat{T}}\vect{x}_{T-\hat{T}}}_2
 \le  \frac{\delta_{2K}}{1-\delta_K} \norm {
\vect{x}_{T-\hat{T}}}_2.
\end{eqnarray}
Similarly, the third term is bounded using Propositions \ref{prop1},
and we obtain
\begin{IEEEeqnarray}{rCl}
\nonumber
\norm{\vect{x} - \hat{\vect{x}}_{SP}}_2 & \le &
\left( 1+\frac{\delta_{2K}}{1-\delta_{K}}\right)\norm{ \vect{x}_{T-\hat{T}}}_2 +
\frac{1}{1-\delta_{K}}\norm{ \matr{D}^*_{\hat{T}}\vect{e}}_2 \\
\nonumber & \le & \frac{1}{1-\delta_{3K}}\norm{
\vect{x}_{T-\hat{T}}}_2 + \frac{1}{1-\delta_{3K}}\norm{
\matr{D}^*_{\hat{T}}\vect{e}}_2,
\end{IEEEeqnarray}
where we have replaced $\delta_K$ and $\delta_{2K}$ with
$\delta_{3K}$, thereby bounding the existing expression from above.
Plugging (\ref{eq:temp_bound1}) into this inequality leads to
\begin{IEEEeqnarray}{rCl}
\nonumber
\norm{\vect{x} - \hat{\vect{x}}_{SP}}_2 & \le &
\frac{1}{1-\delta_{3K}}\left(2 + 2 \cdot \frac{6 - 6\delta_{3K} + 4\delta_{3K}^2}
{(1-\delta_{3K})^3} \right)\norm{ \matr{D}^*_{\hat{T}}\vect{e}}_2\\
\nonumber
& = & 2\cdot \frac{7 - 9\delta_{3K} + 7\delta_{3K}^2 - \delta_{3K}^3}
{(1-\delta_{3K})^4}\norm{ \matr{D}^*_{\hat{T}}\vect{e}}_2.
\end{IEEEeqnarray}
Applying the condition $\delta_{3K} \le 0.139$ on this equation
leads to the result in (\ref{eq:SP_error_bound_consts}). \hfill
$\Box$
\bigskip

For practical use we may suggest a simpler term for $\ell^*$. Since
$\norm{\matr{D}_{T_\vect{e}}^*\vect{e}}_2$ is defined by the subset
that gives the maximal correlation with the noise, and it appears in
the denominator of $\ell^*$, it can be replaced with the average
correlation, thus $\ell^* \approx \ceil{\log_2 \left(
\norm{\vect{x}}_2/\sqrt{K}\sigma \right)}$.

Now that we have a bound for the SP algorithm for the adversial
case, we proceed and consider a bound for the random noise case,
which will lead to a near-oracle performance guarantee for the SP
algorithm.
\begin{thm}
\label{thm:SP_oracle_thm} Assume that $\vect{e}$ is a white Gaussian
noise vector with variance $\sigma^2$ and that the columns of
$\matr{D}$ are normalized. If the condition $\delta_{3K}\le 0.139$
holds, then with probability exceeding $1-(\sqrt{\pi (1+a) \log{N}}
\cdot N^a)^{-1}$ we obtain
\begin{equation}
\label{eq:SP_error_bound_near_oracle} \norm{\vect{x} -
\hat{\vect{x}}_{SP}}_2^2 \le C_{SP}^2 \cdot (2(1+a)\log{N})\cdot K
\sigma^2.
\end{equation}
\end{thm}

{\em Proof:} Following Section 3 in \cite{Candes07Dantzig} it holds
true that
${\bf P}\left(\sup_i\abs{\matr{D}_i^*\vect{e}} > \sigma\cdot
\sqrt{2(1+a)\log{N}}\right) \le 1-(\sqrt{\pi (1+a) \log{N}}\cdot
N^a)^{-1}.$ Combining this with
(\ref{eq:SP_error_bound_consts}), and bearing in mind that
$|T_e|=K$, we get the stated result. \hfill $\Box$
\bigskip

As can be seen, this result is similar to the one posed in
\cite{Candes07Dantzig} for the Dantzig-Selector, but with a
different constant -- the one corresponding to DS is $\approx 5.5$
for the RIP requirement used for the SP. For both algorithms,
smaller values of $\delta_{3K}$ provide smaller constants.


\subsection{Near oracle performance of the CoSaMP algorithm}
\label{sec:CoSaMP_perf}

We continue with the CoSaMP pursuit method, as described in
Algorithm~\ref{alg:CoSaMP}. CoSaMP, in a similar way to the SP,
holds a temporal solution with $K$ non-zero entries, with the
difference that in each iteration it adds an additional set of $2K$
(instead of $K$) candidate non-zeros that are most correlated with
the residual. Anther difference is that after the punning step in SP
we use a matrix inversion in order to calculate a new projection for
the $K$ dominant elements, while in the CoSaMP we just take the
biggest $K$ elements. Here also, we use a constant number of
iterations as a stopping criterion while different stopping criteria
can be sought, as presented in \cite{Needell09CoSaMP}.

\begin{algorithm}
\caption{CoSaMP Algorithm [Algorithm 2.1 in
\cite{Needell09CoSaMP}]} \label{alg:CoSaMP}
\begin{algorithmic}

\REQUIRE $K, \matr{D}, \vect{y}$ where $\vect{y} = \matr{D}\vect{x}
+ \vect{e}$, $K$ is the cardinality of $\vect{x}$ and $\vect{e}$ is
the additive noise.

\ENSURE $\hat{\vect{x}}_{CoSaMP}$: $K$-sparse approximation of
$\vect{x}$

\STATE Initialize the support: $T^0 =\emptyset$.

\STATE Initialize the residual: $\vect{y}_r^0 = \vect{y}$.

\WHILE{halting criterion is not satisfied}

\STATE Find new support elements: $T_\Delta =
\supp(\matr{D}^*\vect{y}^{\ell - 1}_r,2K)$.

\STATE Update the support: $\tilde{T^\ell} = T^{\ell -1} \cup
T_\Delta$.

\STATE Compute the representation: $\vect{x}_p =
\matr{D}^{\dag}_{\tilde{T}^\ell}\vect{y}$.

\STATE Prune small entries in the representation: $T^\ell =
\supp(\vect{x}_p,K)$.

\STATE Update the residual: $\vect{y}_r^\ell = \vect{y} - \matr{D}_{T^\ell}(\vect{x}_p)_{T^\ell}
$.

\ENDWHILE \STATE Form the final solution:
$\hat{\vect{x}}_{CoSaMP,(T^\ell)^C} = 0$ and
$\hat{\vect{x}}_{CoSaMP,T^\ell} = (\vect{x}_p)_{T^\ell}$.
\end{algorithmic}
\end{algorithm}

In the analysis of the CoSaMP that comes next, we follow the same
steps as for the SP to derive a near-oracle performance guarantee.
Since the proofs are very similar to those of the SP, and those
found in \cite{Needell09CoSaMP}, we omit most of the derivations and
present only the differences.

\begin{thm} The CoSaMP solution at the $\ell$-th iteration
satisfies the recurrence inequality\footnote{The observant reader
will notice a delicate difference in terminology between this
theorem and Theorem \ref{thm:SP-1}. While here the recurrence
formula is expressed with respect to the estimation error,
$\norm{\vect{x}-\hat{\vect{x}}_{CoSaMP}^\ell}_2$, Theorem
\ref{thm:SP-1} uses a slightly different error measure,
$\norm{\vect{x}_{T-T^\ell}}_2$.}
\begin{IEEEeqnarray}{rCl} 
\label{eq:CoSaMP_x_diff_bound}
\norm{\vect{x}-\hat{\vect{x}}_{CoSaMP}^\ell}_2 &\le &
\frac{4\delta_{4K}}{(1-\delta_{4K})^2}\norm{\vect{x}-\hat{\vect{x}}_{CoSaMP}^{\ell-1}}_2
\\ \nonumber && + \frac{14-6\delta_{4K} }{(1-\delta_{4K})^2}
\norm{\matr{D}_{T_{\vect{e}}}^*\vect{e}}_2
\end{IEEEeqnarray}
For $\delta_{4K} \le 0.1$ this leads to
\begin{equation}
\label{eq:CoSaMP_x_diff_bound_consts} \norm{\vect{x}-\hat{\vect{x}}_{CoSaMP}^\ell}_2 \le
0.5\norm{\vect{x}-\hat{\vect{x}}_{CoSaMP}^{\ell-1}}_2+16.6\norm{\matr{D}_{T_\vect{e}}^*\vect{e}}_2.
\end{equation}
\end{thm}

{\em Proof:} The proof of the inequality in
(\ref{eq:CoSaMP_x_diff_bound}) is given in Appendix
\ref{sec:CoSaMP_x_diff_bound_proof}.  In a similar way to the proof
in the SP case, under the condition $\delta_{4K} \le 0.1$, it holds
that the coefficient multiplying
$\norm{\vect{x}-\hat{\vect{x}}_{CoSaMP}^{\ell-1}}_2$ is smaller or
equal to $0.5$, while the coefficient multiplying
$\norm{\matr{D}_{T_\vect{e}}^*\vect{e}}_2$ is smaller or equal to
$16.6$, which completes our proof. \hfill $\Box$
\bigskip

\begin{cor}
\label{cor:CoSaMP_bound_cor} Under the condition $\delta_{4K} \le
0.1$, the CoSaMP algorithm satisfies
\begin{equation}
\label{eq:CoSaMP_bound_consts} \norm{\vect{x}-\hat{\vect{x}}_{CoSaMP}^\ell}_2 \le
2^{-\ell}\norm{\vect{x}}_2+2\cdot
16.6\norm{\matr{D}_{T_\vect{e}}^*\vect{e}}_2.
\end{equation}
In addition, After at most
\begin{equation}\label{eq:Num_of_Iter_CoSaMP}
\ell^* = \ceil{\log_2 \left(
\frac{\norm{\vect{x}}_2}{\norm{\matr{D}_{T_\vect{e}}^*\vect{e}}_2}
\right)}
\end{equation}
iterations, the solution $\hat{\vect{x}}_{CoSaMP}$ leads to an accuracy
\begin{equation}
\label{eq:CoSaMP_error_bound_consts} \norm{\vect{x}-\hat{\vect{x}}_{CoSaMP}^\ell}_2 \le
C_{CoSaMP}\norm{\matr{D}_{T_\vect{e}}^*\vect{e}}_2,
\end{equation}
where
\begin{equation}
C_{CoSaMP} = \frac{29-14\delta_{4K} + \delta_{4K}^2
}{(1-\delta_{4K})^2} \le 34.1.
\end{equation}
\end{cor}

{\em Proof:} Starting with (\ref{eq:CoSaMP_x_diff_bound_consts}),
and applying it recursively, in the same way as was done in the
proof of  Corollary \ref{cor:CoSaMP_bound_cor}, we obtain
\begin{eqnarray}
\norm{\vect{x}-\hat{\vect{x}}_{CoSaMP}^\ell}_2 & \le &
0.5^k\norm{\vect{x}-\hat{\vect{x}}_{CoSaMP}^{\ell-k}}_2 \\ \nonumber
&&+ 16.6\cdot \left(\sum_{j=0}^{k-1}
0.5^j\right)\norm{\matr{D}_{T_\vect{e}}^*\vect{e}}_2.
\end{eqnarray}
Setting $k=\ell$ leads easily to (\ref{eq:CoSaMP_bound_consts}), since
$\norm{\vect{x}-\hat{\vect{x}}_{CoSaMP}^0}_2 = \norm{\vect{x}}_2$.

Plugging the number of iterations $\ell^*$ as in
(\ref{eq:Num_of_Iter_CoSaMP}) to
(\ref{eq:CoSaMP_bound_consts}) yields\footnote{As before, we replace
the constant $16.6$ with the equivalent expression that depends on
$\delta_{4K}$ -- see  (\ref{eq:CoSaMP_x_diff_bound}).}
\begin{IEEEeqnarray}{rcl}
\nonumber
\norm{\vect{x}-\hat{\vect{x}}_{CoSaMP}^\ell}_2
& \le & 2^{-\ell^*}\norm{\vect{x}}_2+2\cdot \frac{14-6\delta_{4K}
}{(1-\delta_{4K})^2}\norm{\matr{D}_{T_\vect{e}}^*\vect{e}}_2
\\ \nonumber ~~~~~& \le & \left(1+2\cdot
\frac{14-6\delta_{4K} }{(1-\delta_{4K})^2}\right)
\norm{\matr{D}_{T_\vect{e}}^*\vect{e}}_2
\\ \nonumber ~~~~~& \le & \frac{29-14\delta_{4K}+ \delta_{4K}^2 }
{(1-\delta_{4K})^2}\norm{\matr{D}_{T_\vect{e}}^*\vect{e}}_2.
\end{IEEEeqnarray}
Applying the condition $\delta_{4K} \le 0.1$ on this equation
leads to the result in (\ref{eq:CoSaMP_error_bound_consts}). \hfill $\Box$
\bigskip

As for the SP, we move now to the random noise case, which leads to
a near-oracle performance guarantee for the CoSaMP algorithm.
\begin{thm}
\label{thm:CoSaMP_oracle_thm} Assume that $\vect{e}$ is a white
Gaussian noise vector with variance $\sigma^2$ and that the columns
of $\matr{D}$ are normalized. If the condition $\delta_{4K}\le 0.1$
holds, then with probability exceeding $1-(\sqrt{\pi (1+a) \log{N}}
\cdot N^a)^{-1}$ we obtain
\begin{equation}
\label{eq:CoSaMP_error_bound_near_oracle} \norm{\vect{x} -
\hat{\vect{x}}_{CoSaMP}}_2^2 \le C_{CoSaMP}^2 \cdot (2(1+a)\log{N})\cdot K
\sigma^2.
\end{equation}
\end{thm}

{\em Proof:} The proof is identical  to the one of Theorem
\ref{thm:CoSaMP_oracle_thm}. \hfill $\Box$
\bigskip

\begin{figure}[!t]
\begin{center}
\includegraphics[width=0.45\textwidth]{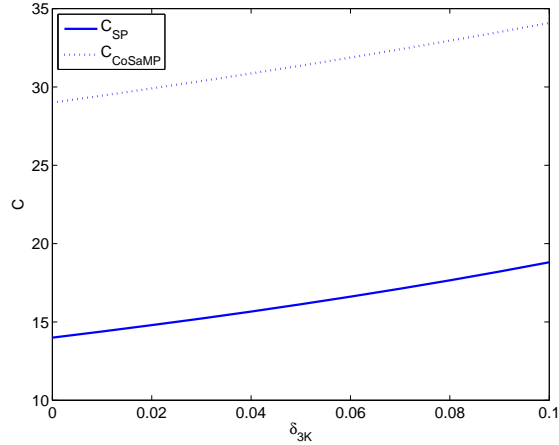}
\end{center}
\caption{The constants of the SP and CoSaMP algorithms as a funtion
of $\delta_{4K}$} \label{fig:C_SP_CoSaMP}
\end{figure}

Fig.~\ref{fig:C_SP_CoSaMP} shows a graph of $C_{CoSaMP}$ as a
function of $\delta_{4K}$. In order to compare the CoSaMP to SP, we
also introduce in this figure a graph of $C_{SP}$ versus
$\delta_{4K}$ (replacing $\delta_{3K}$). Since $\delta_{3K} \le
\delta_{4K}$, the constant $C_{SP}$ is actually better than the
values shown in the graph, and yet, it can be seen that even in this
case we get $C_{SP}<C_{CoSaMP}$. In addition, the requirement for
the SP is expressed with respect to $\delta_{3K}$, while the
requirement for the CoSaMP is stronger and uses $\delta_{4K}$.

With comparison to the results presented in
\cite{BenHaim09Coherence} for the OMP and the thresholding, the
results obtained for the CoSaMP and SP are uniform, expressed only
with respect to the properties of the dictionary $\matr{D}$. These
algorithms' validity is not dependent on the values of the input
vector $\vect{x}$, its energy, or the noise power. The condition
used is the RIP, which implies constraints only on the used
dictionary and the sparsity level.


\subsection{Near oracle performance of the IHT algorithm}
\label{sec:IHT_perf}

The IHT algorithm,  presented in Algorithm \ref{alg:IHT}, uses a
different strategy than the SP and the CoSaMP. It applies only
multiplications by $\matr{D}$ and $\matr{D}^*$, and a hard
thresholding operator. In each iteration it calculates a new
representation and keeps its $K$ largest elements. As for the SP and
CoSaMP, here as well we employ a fixed number of iterations as a
stopping criterion.

\begin{algorithm}
\caption{IHT Algorithm [Equation 7 in \cite{Blumensath09Iterative}]} \label{alg:IHT}
\begin{algorithmic}

\REQUIRE $K, \matr{D}, \vect{y}$ where $\vect{y} = \matr{D}\vect{x}
+ \vect{e}$, $K$ is the cardinality of $\vect{x}$ and $\vect{e}$ is
the additive noise.

\ENSURE $\hat{\vect{x}}_{IHT}$: $K$-sparse approximation of
$\vect{x}$

\STATE Initialize the support: $T^0 =\emptyset$.

\STATE Initialize the representation: $\vect{x}^0_{IHT} = 0$.

\WHILE{halting criterion is not satisfied}

\STATE Compute the representation: $\vect{x}_p =
\hat{\vect{x}}_{IHT}^{\ell-1} + \matr{D}^*(\vect{y} -
\matr{D}\hat{\vect{x}}_{IHT}^{\ell-1})$.

\STATE Prune small entries in the representation: $T^\ell =
\supp(\vect{x}_p,K)$.

\STATE Update the representation: $\hat{\vect{x}}_{IHT,(T^\ell)^C}^{\ell} = 0$ and
$\hat{\vect{x}}_{IHT,T^\ell}^{\ell} = (\vect{x}_p)_{T^\ell}$.

\ENDWHILE \STATE Form the final solution:
$\hat{\vect{x}}_{IHT,(T^\ell)^C} = 0$ and
$\hat{\vect{x}}_{IHT,T^\ell} = (\vect{x}_p)_{T^\ell}$.
\end{algorithmic}
\end{algorithm}

Similar results, as of the SP and CoSaMP methods, can be sought for
the IHT method. Again, the proofs are very similar to the ones shown
before for the SP and the CoSaMP and thus only the differences will
presented.

\begin{thm} The IHT solution at the $\ell$-th iteration
satisfies the recurrence inequality
\begin{IEEEeqnarray}{c} 
\label{eq:IHT_x_diff_bound} \norm{\vect{x} -
\hat{\vect{x}}^\ell_{IHT}}_2 \le
\sqrt{8}\delta_{3K}\norm{\vect{x}-\hat{\vect{x}}_{IHT}^{\ell-1}}_2  + 4\norm{
\matr{D}^*_{T_{\vect{e}}}\vect{e}}_2.
\end{IEEEeqnarray}
For $\delta_{3K} \le \frac{1}{\sqrt{32}}$ this leads to
\begin{IEEEeqnarray}{c} 
\label{eq:IHT_x_diff_bound_consts}  \norm{\vect{x} -
\hat{\vect{x}}^\ell_{IHT}}_2 \le
0.5\norm{\vect{x}-\hat{\vect{x}}_{IHT}^{\ell-1}}_2  + 4\norm{
\matr{D}^*_{T_{\vect{e}}}\vect{e}}_2.
\end{IEEEeqnarray}
\end{thm}

{\em Proof:} Our proof is based on the proof of Theorem 5 in
\cite{Blumensath09Iterative}, and only major modifications in the
proof will be presented here. Using the definition $\vect{r}^\ell
\triangleq \vect{x}-\hat{\vect{x}}_{IHT}^\ell$, and an inequality
taken from Equation (22) in \cite{Blumensath09Iterative}, it holds
that
\begin{IEEEeqnarray}{rcl}
\label{eq:IHT_proof_step1}
&& \norm{\vect{x} - \hat{\vect{x}}^\ell_{IHT}}_2 \le
2 \norm{\vect{x}_{T\cup T^\ell} - (\vect{x}_p)_{T\cup T^\ell}}_2 \\
 \nonumber && ~~~~= 2 \norm{\vect{x}_{T\cup T^\ell} -
\hat{\vect{x}}^{\ell-1}_{T\cup T^\ell} -\matr{D}^*_{T\cup
T^\ell}\matr{D}\vect{r}^{\ell-1} - \matr{D}^*_{T\cup
T^\ell}\vect{e}}_2
\\ \nonumber && ~~~~ = 2 \norm{\vect{r}_{T\cup
T^\ell}^{\ell-1} -\matr{D}^*_{T\cup T^\ell}\matr{D}\vect{r}^{\ell-1}
- \matr{D}^*_{T\cup T^\ell}\vect{e}}_2,
\end{IEEEeqnarray}
where the equality emerges from the definition $\vect{x}_p =
\hat{\vect{x}}_{IHT}^{\ell-1} + \matr{D}^*(\vect{y} -
\matr{D}\hat{\vect{x}}_{IHT}^{\ell-1}) =
\hat{\vect{x}}_{IHT}^{\ell-1} + \matr{D}^*(\matr{D}\vect{x} +
\vect{e}- \matr{D}\hat{\vect{x}}_{IHT}^{\ell-1}).$

The support of $\vect{r}^{\ell-1}$ is over $T\cup T^{\ell-1} $ and
thus it is also over $T \cup T^\ell \cup T^{\ell -1}$. Based on
this, we can divide $\matr{D}\vect{r}^{\ell - 1}$ into a part
supported on $T^{\ell-1}- T^\ell \cup T$ and a second part supported
on $T^\ell \cup T$. Using this and the triangle inequality with
(\ref{eq:IHT_proof_step1}), we obtain
\begin{IEEEeqnarray}{rCl}
\label{eq:IHT_proof_step2}
&&\norm{\vect{x} - \hat{\vect{x}}^\ell_{IHT}}_2 \\ \nonumber && ~~ \le  2
\norm{\vect{r}^{\ell-1}_{T\cup T^\ell} -
\matr{D}^*_{T\cup T^\ell}\matr{D}\vect{r}^{\ell-1}}_2 +
2\norm{ \matr{D}^*_{T\cup T^\ell}\vect{e}}_2 \\
\nonumber  && ~~ =  2\left\Vert (\matr{I} - \matr{D}^*_{T\cup
T^\ell}\matr{D}_{T\cup T^\ell}) \vect{r}^{\ell-1}_{T\cup T^\ell} \right. \\ \nonumber &&
~~~~\left. -\matr{D}^*_{T\cup T^\ell}\matr{D}_{T^{\ell-1} - T\cup
T^\ell}\vect{r}^{\ell-1}_{T^{\ell-1} - T\cup T^\ell} \right\Vert _2
 +2\norm{ \matr{D}^*_{T\cup T^\ell}\vect{e}}_2 \\
\nonumber && ~~
 \le  2\norm{(\matr{I} - \matr{D}^*_{T\cup T^\ell}\matr{D}_{T\cup T^\ell})
\vect{r}^{\ell-1}_{T\cup T^\ell}}_2 \\ \nonumber &&  ~~~~
 + 2\norm{  \matr{D}^*_{T\cup T^\ell}
\matr{D}_{T^{\ell-1} - T\cup T^\ell}\vect{r}^{\ell-1}_{T^{\ell-1} - T\cup T^\ell}}_2
 + 2\norm{ \matr{D}^*_{T_{\vect{e}},2}\vect{e}}_2 \\
\nonumber && ~~ \le  2\delta_{2K}\norm{\vect{r}^{\ell-1}_{T\cup
T^\ell}}_2 + 2\delta_{3K}\norm{\vect{r}^{\ell-1}_{T^{\ell-1} - T\cup
T^\ell}}_2 + 4\norm{ \matr{D}^*_{T_{\vect{e}}}\vect{e}}_2.
\end{IEEEeqnarray}
The last inequality holds because the eigenvalues of $(\matr{I} -
\matr{D}^*_{T\cup T^\ell}\matr{D}_{T\cup T^\ell})$ are in the range
$[-\delta_{2K},\delta_{2K}]$, the size of the set $T\cup T^\ell$ is
smaller than $2K$, the sets $T\cup T^\ell$ and $T^{\ell-1} - T\cup
T^\ell$ are disjoint, and of total size of these together is equal
or smaller than $3K$. Note that we have used the definition of
$T_{\vect{e},2}$ as given in Section \ref{sec:notation}.

We proceed by observing that $\norm{\vect{r}^{\ell-1}_{T^{\ell-1} -
T\cup T^\ell}}_2 + \norm{\vect{r}^{\ell-1}_{T\cup T^\ell}}_2 \le
\sqrt{2}\norm{\vect{r}^{\ell-1}}_2$,  since these vectors are
orthogonal. Using the fact that $\delta_{2K} \le \delta_{3K}$ we get
(\ref{eq:IHT_x_diff_bound}) from (\ref{eq:IHT_proof_step2}).
Finally, under the condition $\delta_{3K} \le 1/\sqrt{32}$, it holds
that the coefficient multiplying
$\norm{\vect{x}-\hat{\vect{x}}_{IHT}^{\ell-1}}_2$ is smaller or
equal to $0.5$, which completes our proof. \hfill $\Box$
\bigskip

\begin{cor}
\label{cor:IHT_bound_cor} Under the condition $\delta_{3K} \le
1/\sqrt{32}$, the IHT algorithm satisfies
\begin{equation}
\label{eq:IHT_bound_consts} \norm{\vect{x}-\hat{\vect{x}}_{IHT}^\ell}_2 \le
2^{-\ell}\norm{\vect{x}}_2+
8\norm{\matr{D}_{T_\vect{e}}^*\vect{e}}_2.
\end{equation}
In addition, After at most
\begin{equation}\label{eq:Num_of_Iter_IHT}
\ell^* = \ceil{\log_2 \left(
\frac{\norm{\vect{x}}_2}{\norm{\matr{D}_{T_\vect{e}}^*\vect{e}}_2}
\right)}
\end{equation}
iterations, the solution $\hat{\vect{x}}_{IHT}$ leads to an accuracy
\begin{equation}
\label{eq:IHT_error_bound_consts} \norm{\vect{x}-\hat{\vect{x}}_{IHT}^\ell}_2 \le
C_{IHT}\norm{\matr{D}_{T_\vect{e}}^*\vect{e}}_2,
\end{equation}
where
\begin{equation}
C_{IHT} = 9.
\end{equation}
\end{cor}

{\em Proof:} The proof is obtained following the same steps as in
Corollaries \ref{cor:SP_bound_cor} and \ref{cor:CoSaMP_bound_cor}.
\hfill $\Box$
\bigskip

Finally, considering a random noise instead of an adversial one, we
get a near-oracle performance guarantee for the IHT algorithm, as
was achieved for the SP and CoSaMP.
\begin{thm}
\label{thm:IHT_oracle_thm} Assume that $\vect{e}$ is a white
Gaussian noise with variance $\sigma^2$ and that the columns of
$\matr{D}$ are normalized. If the condition $\delta_{3K}\le
1/\sqrt{32}$ holds, then with probability exceeding $1-(\sqrt{\pi
(1+a) \log{N}} \cdot N^a)^{-1}$ we obtain
\begin{equation}
\label{eq:IHT_error_bound_near_oracle} \norm{\vect{x} -
\hat{\vect{x}}_{IHT}}_2^2 \le C_{IHT}^2 \cdot (2(1+a)\log{N})\cdot K
\sigma^2.
\end{equation}
\end{thm}

{\em Proof:} The proof is identical to the one of  Theorem
\ref{thm:CoSaMP_oracle_thm}. \hfill $\Box$
\bigskip

\begin{figure}[!t]
\begin{center}
\includegraphics[width=0.45\textwidth]{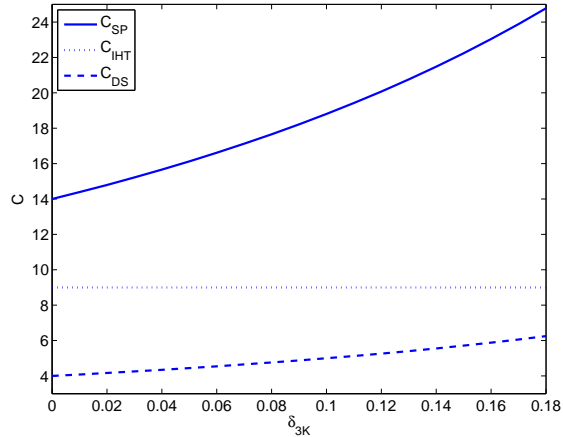}
\end{center}
\caption{The constants of the SP, IHT and DS algorithms as a funtion
of $\delta_{3K}$} \label{fig:C_SP_IHT_DS}
\end{figure}

A comparison between the constants achieved by the IHT, SP and DS is
presented in Fig.~\ref{fig:C_SP_IHT_DS}. The CoSaMP constant was
omitted since it is bigger than the one of the SP and it is
dependent on  $\delta_{4K}$ instead of $\delta_{3K}$. The figure
shows that the constant values of IHT and DS are better than that of
the SP (and as such better than the one of the CoSaMP), and that the
one of the DS is the smallest. It is interesting to note that the
constant of the IHT is independent of $\delta_{3K}$.

In table \ref{tbl:comparison} we summarize the performance
guarantees of several different algorithms -- the DS
\cite{Candes07Dantzig}, the BP \cite{Bickel09Simultaneous}, and the
three algorithms analyzed in this paper.

\begin{table*}[!t]
\begin{center}\small{
\begin{tabular}{|l|l|l|l|l|}
\hline Alg. & RIP Condition & Probability of Correctness
& Constant & The Obtained Bound \\
\hline DS & $\delta_{2K} + \delta_{3K} \le 1$ & $1-(\sqrt{\pi
(1+a)\log{N}}\cdot N^a)^{-1}$ & $ \frac{4}{1-2\delta_{3K}} $ &
$C_{DS}^2 \cdot (2(1+a)\log{N})\cdot K  \sigma^2$ \\ \hline BP  &
$\delta_{2K} + 3\delta_{3K} \le 1$ & $1-(N^a)^{-1}$ &
$>\frac{32}{\kappa^4}$ & $C_{BP}^2 \cdot (2(1+a)\log{N})\cdot K
\sigma^2$ \\ \hline SP & $\delta_{3K}\le 0.139$ & $ 1-(\sqrt{\pi
(1+a)\log{N}}\cdot N^a)^{-1}$ &$ \le  21.41$ & $C_{SP}^2 \cdot
(2(1+a)\log{N})\cdot K \sigma^2$ \\ \hline CoSaMP  & $\delta_{4K}
\le 0.1$ & $ 1-(\sqrt{\pi (1+a)\log{N}}\cdot N^a)^{-1}$ & $  \le
34.2$ & $C_{CoSaMP}^2 \cdot (2(1+a)\log{N})\cdot K \sigma^2$ \\
\hline IHT  & $\delta_{3K} \le \frac{1}{\sqrt{32}}$ & $ 1-(\sqrt{\pi
(1+a)\log{N}}\cdot N^a)^{-1}$
& $9$ & $C_{IHT}^2 \cdot (2(1+a)\log{N})\cdot K \sigma^2$ \\
\hline
\end{tabular}}
\caption{Near oracle performance guarantees for the DS, BP, SP,
CoSaMP and IHT techniques.} \label{tbl:comparison}
\end{center}
\end{table*}

\noindent We can observe the following:
\begin{enumerate}
\item In terms of the RIP: DS and BP are the best, then IHT, then SP and last is
CoSaMP.
\item In terms of the constants in the bounds: the smallest
constant is achieved by DS. Then come IHT, SP, CoSaMP and BP in this
order.
\item In terms of the probability: all have the same probability except
the BP which gives a weaker guarantee.

\item Though the CoSaMP has a weaker guarantee compared to the SP, it has
an efficient implementation that saves the matrix inversion in the
algorithm.\footnote{The proofs of the guarantees in this paper are
not valid for this case, though it is not hard to extend them for
it.}
\end{enumerate}

For completeness of the discussion here, we also refer to
algorithms' complexity: the IHT is the cheapest, CoSaMP and SP come
next with a similar complexity (with a slight advantage to CoSaMP).
DS and BP seem to be the most complex.


Interestingly, in the guarantees of the OMP and the thresholding in
\cite{BenHaim09Coherence} better constants are obtained. However,
these results, as mentioned before, holds under mutual-coherence
based conditions, which are more restricting. In addition, their
validity relies on the magnitude of the entries of $\vect{x}$ and
the noise power, which is not correct for the results presented in
this section for the greedy-like methods. Furthermore, though we get
bigger constants with these methods, the conditions are not tight,
as will be seen in the next section.


\section{Experiments}
\label{sec:experiments}

In our experiments we use a random dictionary with entries drawn
from the canonic normal distribution. The columns of the dictionary
are normalized and the dimensions are $m=512$ and $N=1024$. The
vector $\vect{x}$ is generated by selecting a support uniformly at
random. Then the elements in the support are generated using the
following model\footnote{This model is taken from the experiments
section in \cite{Candes07Dantzig}.}:
\begin{equation}
\vect{x}_i =  10\epsilon_i(1+ \abs{n_i})
\end{equation}
where $\epsilon_i$ is $\pm 1$ with probability $0.5$, and $n_i$ is a
canonic normal random variable. The support and the non-zero values
are statistically independent. We repeat each experiment $1500$
times.

\begin{figure}[!t]
\centering
\subfigure[SP method]{
\includegraphics[width=0.32\textwidth]{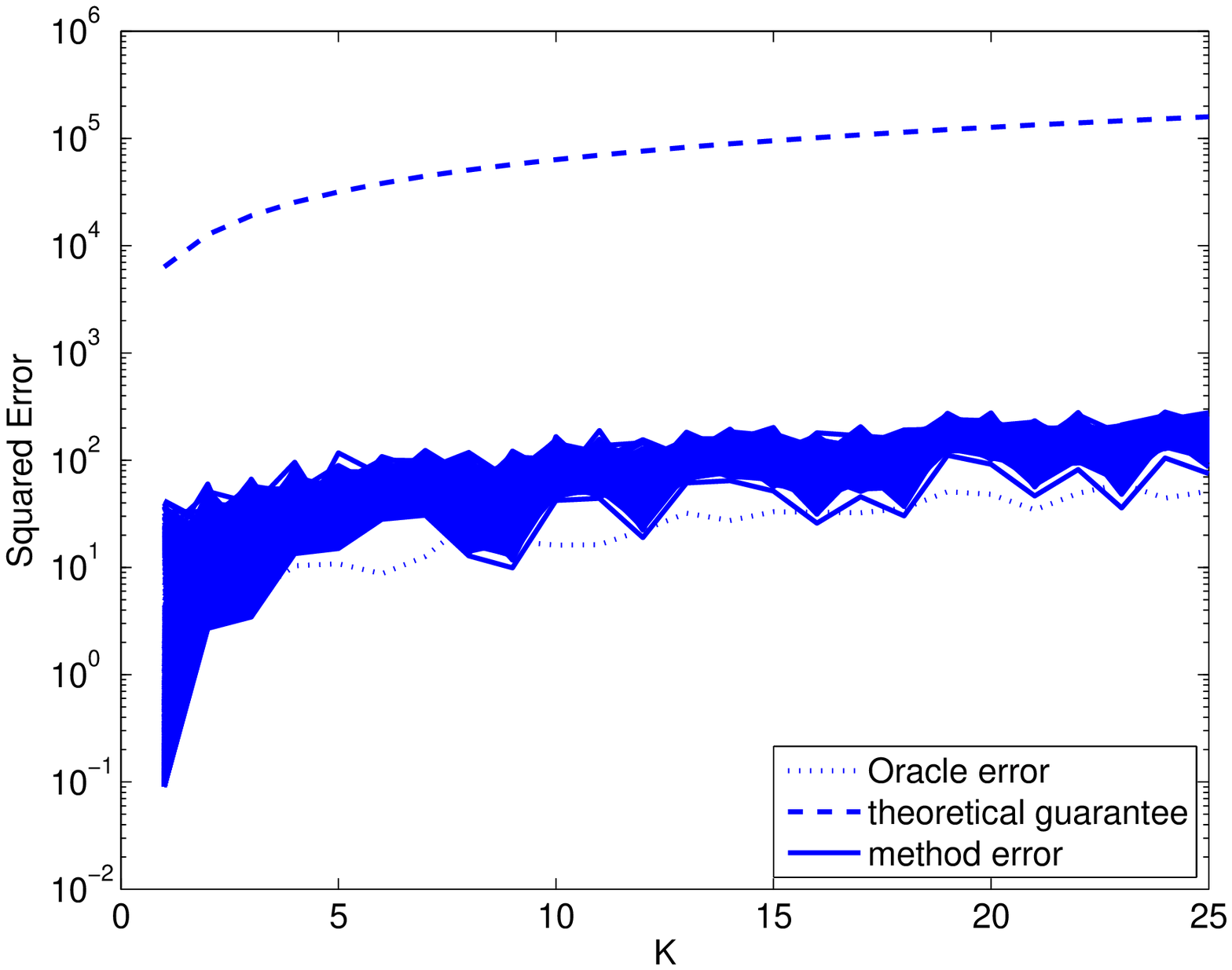}
\label{fig:sp_sparsity_guarantee}
}
\subfigure[CoSaMP method]{
\includegraphics[width=0.32\textwidth]{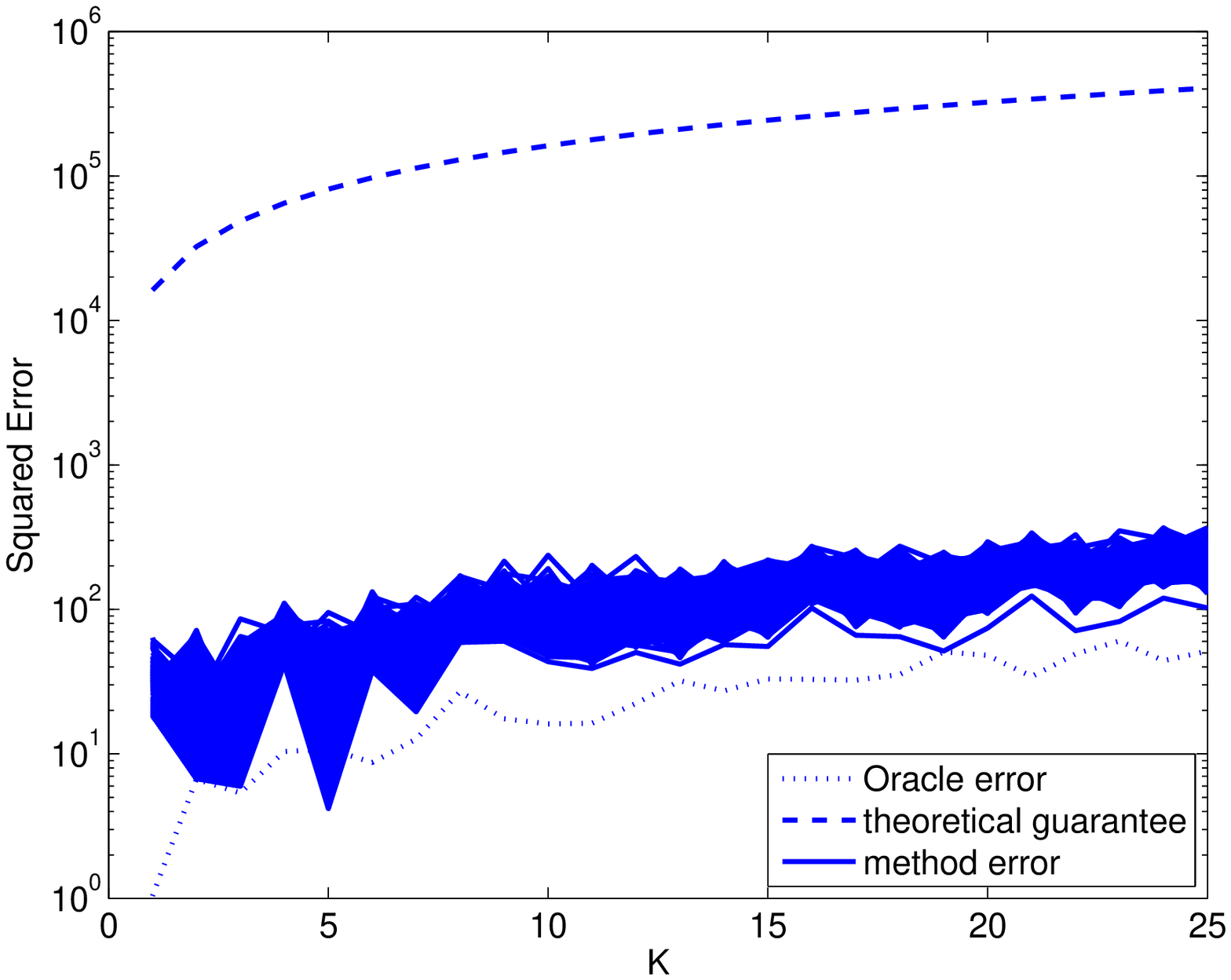}
\label{fig:cosamp_sparsity_guarantee}
}
\subfigure[IHT method]{
\includegraphics[width=0.32\textwidth]{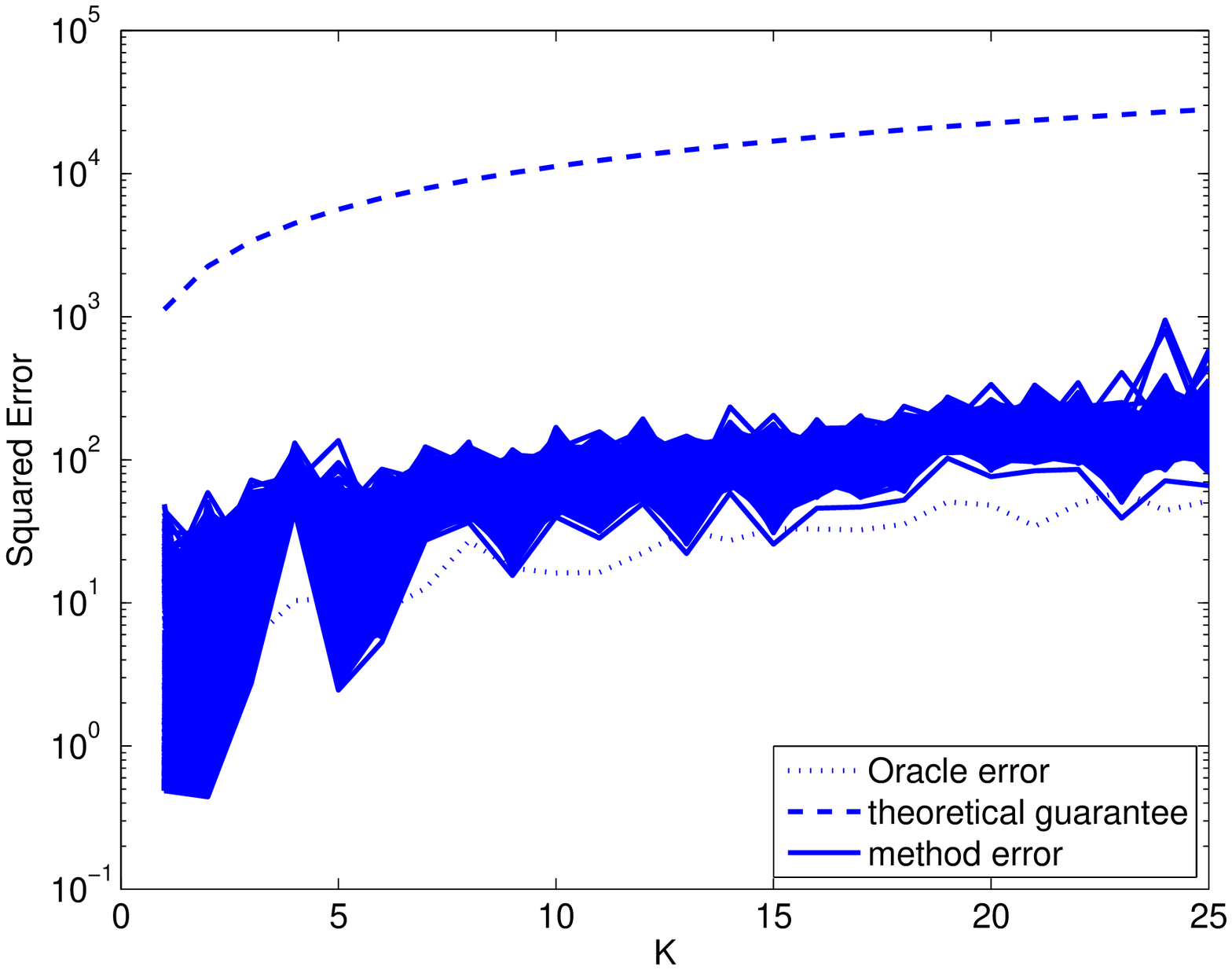}
\label{fig:IHT_sparsity_guarantee} } \caption{The squared-error as
achieved by the SP, the CoSaMP and the IHT algorithms as a function
of the cardinality. The graphs also show the theoretical guarantees
and the oracle performance.}\label{fig:sparsity_guarantee}
\end{figure}

In the first experiment we calculate the error of the SP, CoSaMP and
IHT methods for different sparsity levels. The noise variance is set
to $\sigma = 1$. Fig.~\ref{fig:sparsity_guarantee} presents the
squared-error $\norm{\vect{x}-\vect{\hat x}}_2^2$ of all the
instances of the experiment for the three algorithms. Our goal is to
show that with high-probability the error obtained is bounded by the
guarantees we have developed. For each algorithm we add the
theoretical guarantee and the oracle performance. As can be seen,
the theoretical guarantees are too loose and the actual performance
of the algorithms is much better. However, we see that both the
theoretical and the empirical performance curves show a
proportionality to the oracle error. Note that the actual
performance of the algorithms' may be better than the oracle's --
this happens because the oracle is the Maximum-Likelihood Estimator
(MLE) in this case \cite{BenHaim10Cramer}, and by adding a bias one
can perform even better in some cases.

\begin{figure}[!t]
\centering
\subfigure[RIP condition satisfied]{
\includegraphics[width=0.32\textwidth]{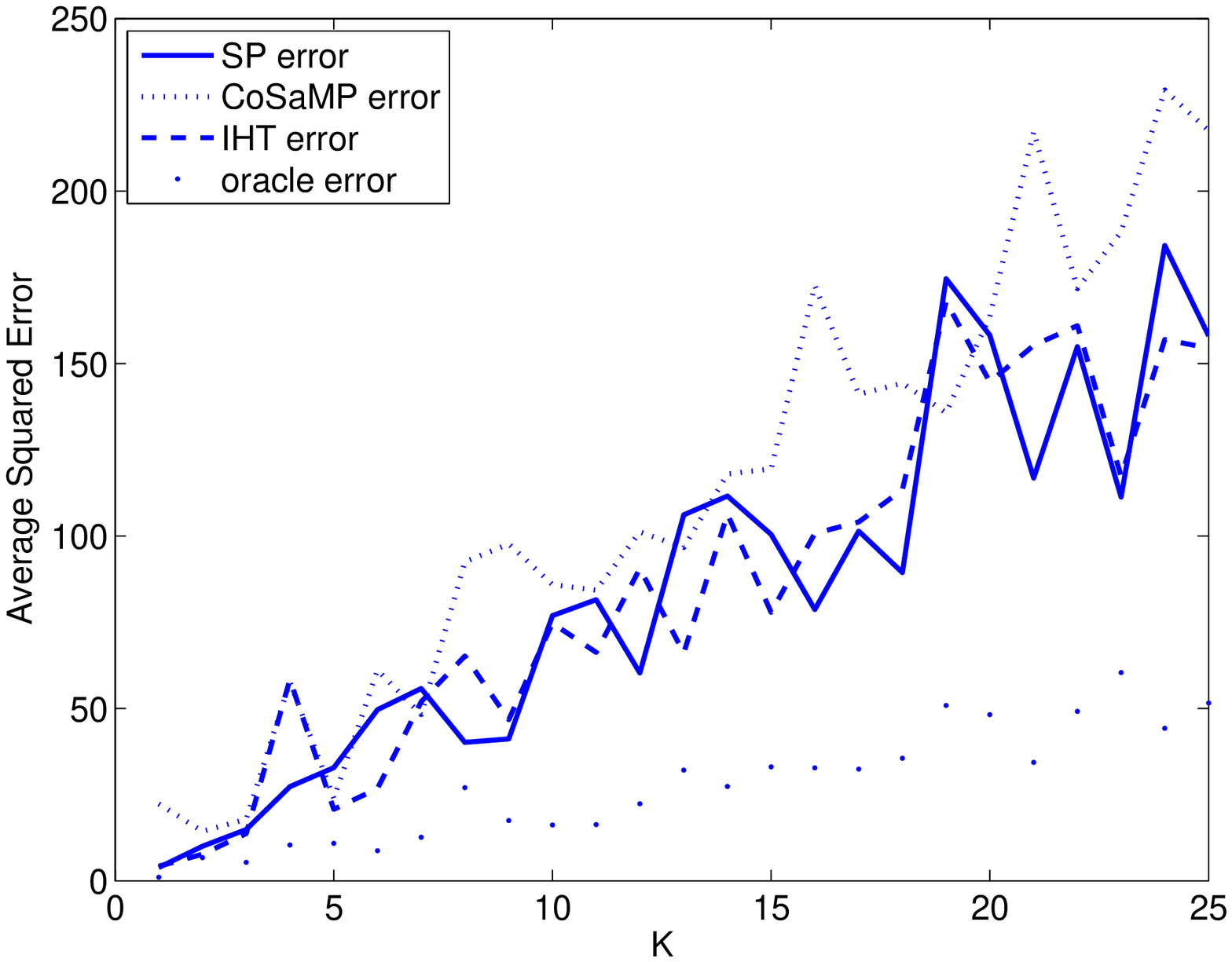}
\label{fig:sparsity_error_range_sp_cosamp_iht} } \subfigure[RIP
condition not satisfied]{
\includegraphics[width=0.32\textwidth]{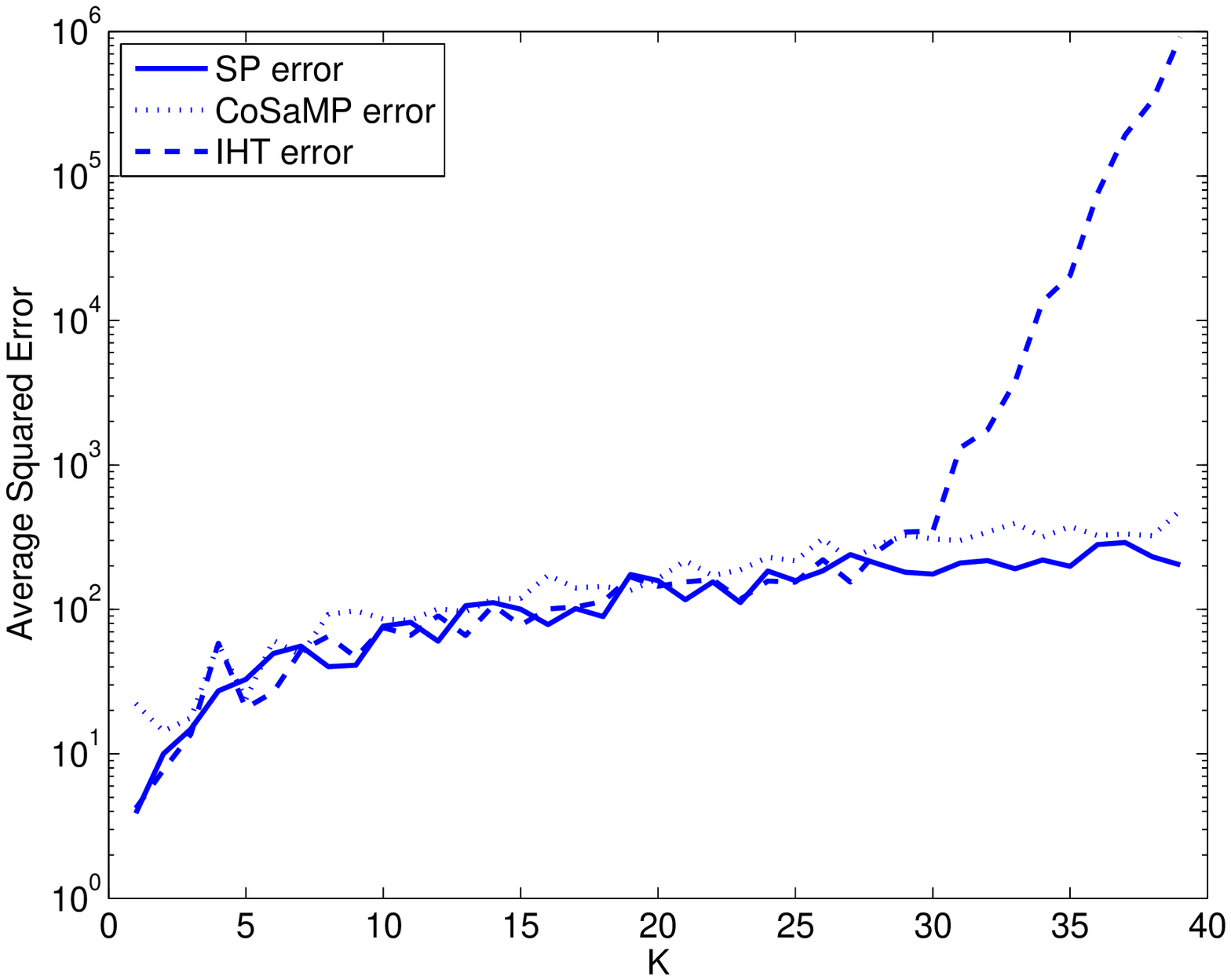}
\label{fig:sparsity_error_wide_sp_cosamp_iht} }
 \caption{The
mean-squared-error of the SP, the CoSaMP and the IHT algorithms as a
function of the
cardinality.}\label{fig:sparsity_error_sp_cosamp_iht}
\end{figure}

Fig.~\ref{fig:sparsity_error_range_sp_cosamp_iht} presents the
mean-squared-error (by averaging all the experiments) for the range
where the RIP-condition seems to hold, and
Fig.~\ref{fig:sparsity_error_wide_sp_cosamp_iht} presents this error for
a wider range, where it is likely top be violated.  It can be seen
that in the average case, though the algorithms get different
constants in their bounds, they achieve almost the same performance.
We also see a near-linear curve describing the error as a function
of $K$. Finally, we observe that the SP and the CoSaMP, which were
shown to have worse constants in theory, have better performance and
are more stable in the case where the RIP-condition do not hold
anymore.

\begin{figure}[!t]
\centering
\subfigure[SP method]{
\includegraphics[width=0.32\textwidth]{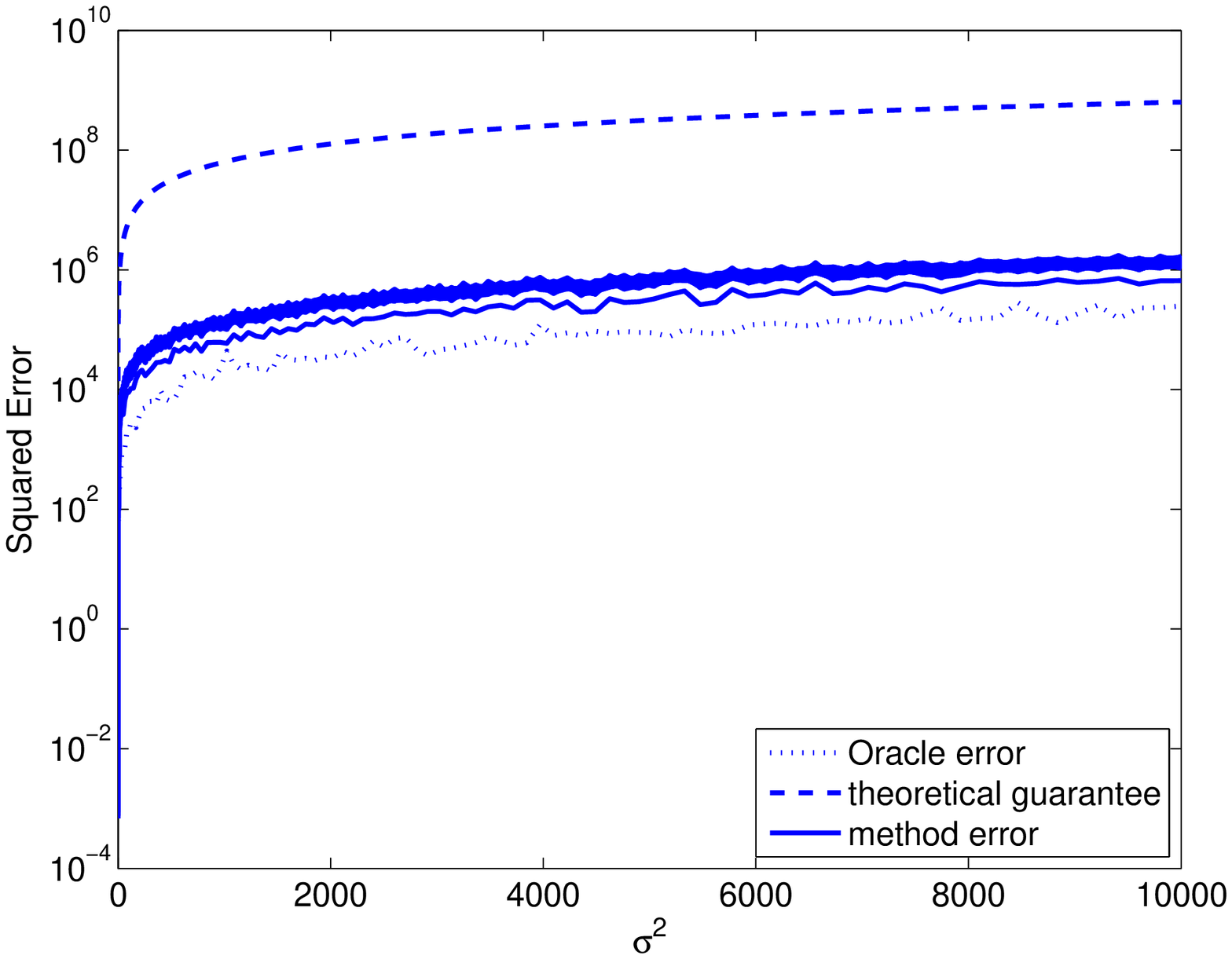}
\label{fig:sp_sigmas_guarantee}
}
\subfigure[CoSaMP method]{
\includegraphics[width=0.32\textwidth]{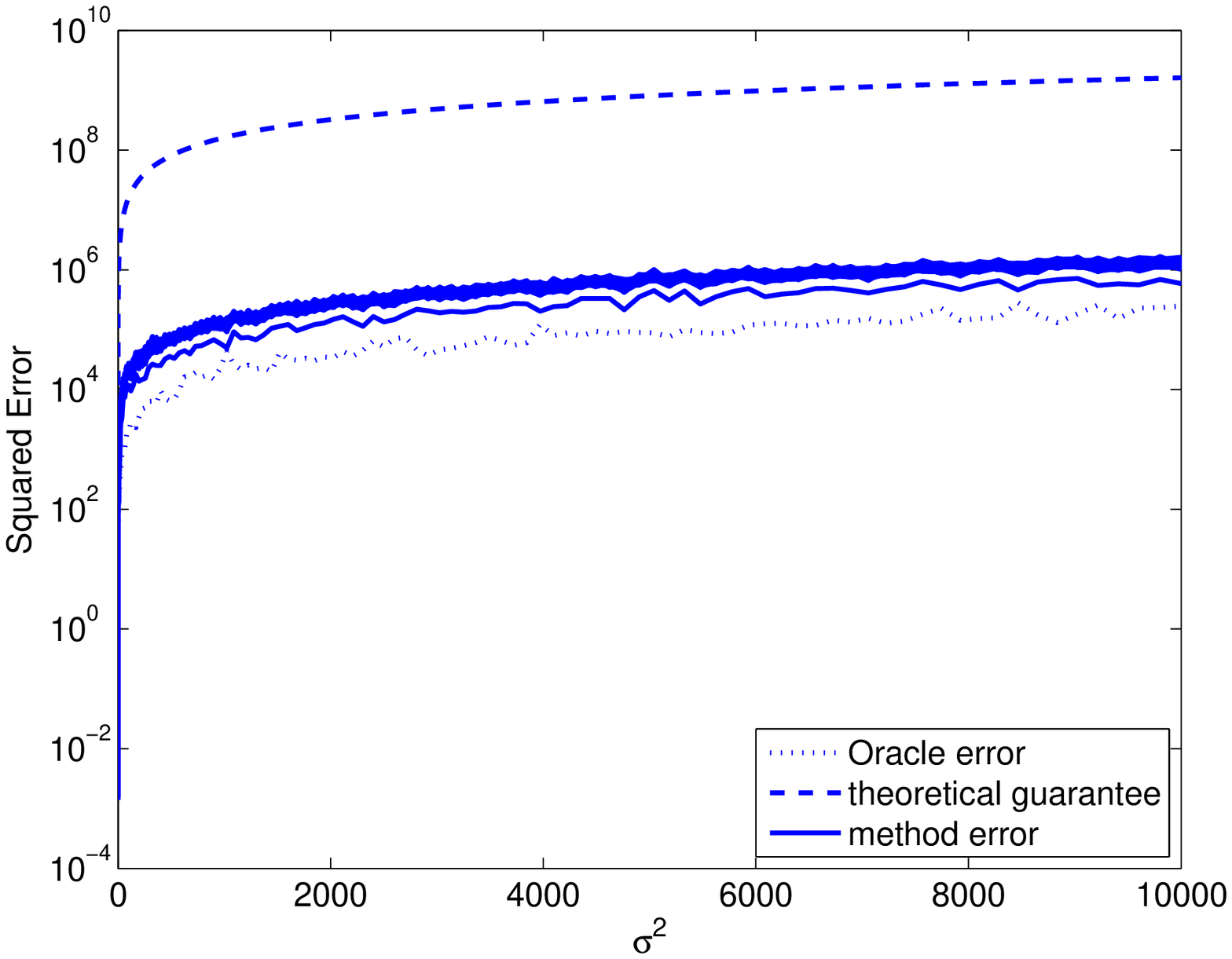}
\label{fig:cosamp_sigma_guarantee}
}
\subfigure[IHT method]{
\includegraphics[width=0.32\textwidth]{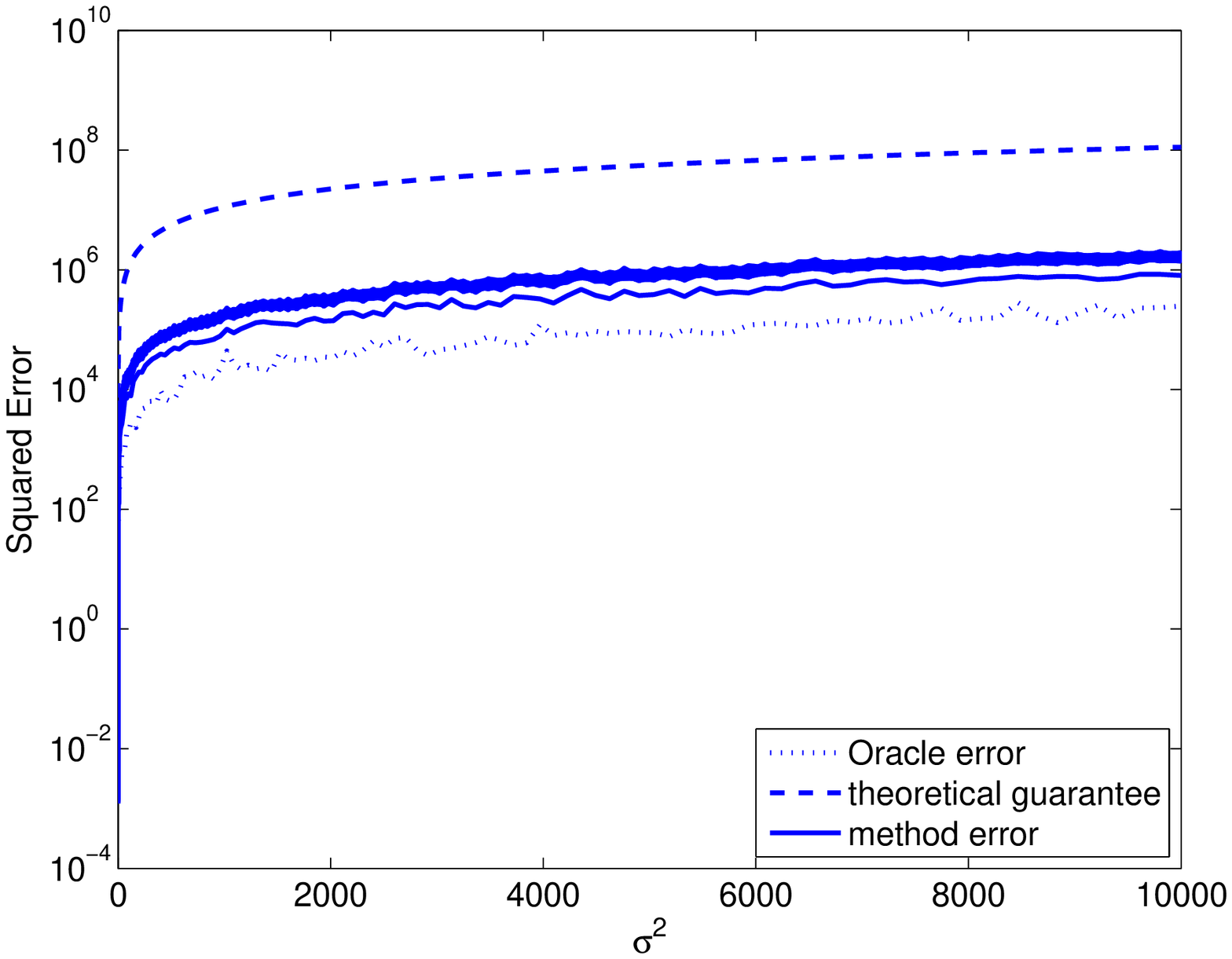}
\label{fig:IHT_sigma_guarantee} } \caption{The squared-error as
achieved by the SP, the CoSaMP and the IHT algorithms as a function
of the noise variance. The graphs also show the theoretical
guarantees and the oracle performance.}\label{fig:sigma_guarantee}
\end{figure}

In a second experiment we calculate the error of the SP, the CoSaMP
and the IHT methods for different noise variances. The sparsity is
set to $K = 10$. Fig.~\ref{fig:sigma_guarantee} presents the error
of all the instances of the experiment for the three algorithms.
Here as well we add the theoretical guarantee and the oracle
performance. As we saw before, the guarantee is not tight but the
error is proportional to the oracle estimator's error.

\begin{figure}[!t]
\begin{center}
\includegraphics[width=0.45\textwidth]{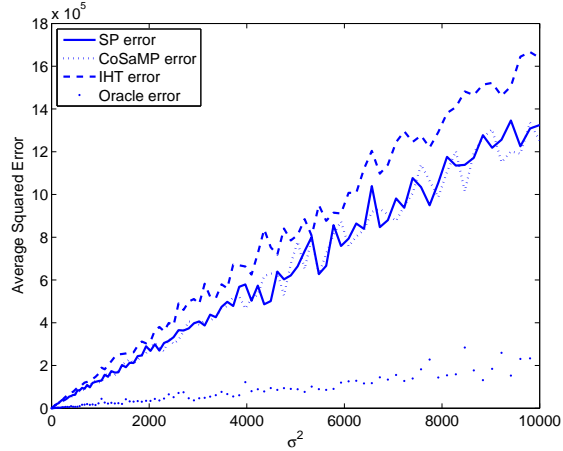}
\end{center}
\caption{The mean-squared-error of the SP, the CoSaMP and the IHT
algorithms as a function of the noise
variance.}\label{fig:power_error_sp_cosamp_iht}
\end{figure}

Fig.~\ref{fig:power_error_sp_cosamp_iht} presents the
mean-squared-error as a function of the noise variance, by averaging
over all the experiments. It can be seen that the error behaves
linearly with respect to the variance, as expected from the
theoretical analysis. Again we see that the constants are not tight
and that the algorithms behave in a similar way. Finally, we note
that the algorithms succeed in meeting the bounds even in very low
signal-to-noise ratios, where simple greedy algorithms are expected
to fail.

\section{Extension to the non-exact sparse case}
\label{sec:non_exact_sparse}

In the case where $\vect{x}$ is not exactly $K$-sparse, our analysis
has to change. Following the work reported in
\cite{Needell09CoSaMP}, we have the following error bounds for all
algorithms (with the different RIP condition and constant).
\begin{thm}
For the SP, CoSaMP and IHT algorithms, under their appropriate RIP
conditions,  it holds that after at most
\begin{equation}
\ell^* = \ceil{\log_2 \left( \frac{\norm{\vect{x}}_2}
{\norm{\matr{D}_{T_\vect{e}}^*\vect{e}}_2} \right)}
\end{equation}
iterations, the estimation $\hat{\vect{x}}$ gives an accuracy of the
form
\begin{IEEEeqnarray}{rCl}
\label{eq:approx_error_bound_consts}
\norm{\vect{x} - \hat{\vect{x}}}_2 &\le& C
\bigg( \norm{\matr{D}_{T_\vect{e}}^*\vect{e}}_2   \\ \nonumber && + (1+\delta_K)
\norm{\vect{x} - \vect{x}_K}_2   + \frac{1+\delta_K}{\sqrt{K}}
\norm{\vect{x} - \vect{x}_K}_1 \bigg) .
\end{IEEEeqnarray}
where $\vect{x}_K$ is a $K$-sparse vector that nulls all entries in
$\vect{x}$ apart from the $K$ dominant ones. $C$ is the appropriate
constant value, dependent on the algorithm.

If we assume that $\vect{e}$ is a white Gaussian noise with variance
$\sigma^2$ and that the columns of $\matr{D}$ are normalized, then
with probability exceeding $1-(\sqrt{\pi (1+a)\log{N}}\cdot N^a)^{-1}$ we
get that
\begin{IEEEeqnarray}{rCl} 
\label{eq:approx_error_bound_near_oracle}
\norm{\vect{x} - \hat{\vect{x}}}_2^2 &\le& 2\cdot C^2
\bigg( \sqrt{(1+a)\log{N}\cdot K} \cdot \sigma  \\ \nonumber && ~~~~~~~~+
\norm{\vect{x} - \vect{x}_K}_2 + \frac{1}{\sqrt{K}}
\norm{\vect{x} - \vect{x}_K}_2 \bigg)^2.
\end{IEEEeqnarray}
\end{thm}

{\em Proof:} Proposition 3.5 from \cite{Needell09CoSaMP} provides us
with the following claim
\begin{eqnarray}
\label{eq:near_approx_bound}
\norm{\matr{D}\vect{x}}_2 \le \sqrt{1+\delta_K}
\left( \norm{\vect{x}}_2 +\frac{1}{\sqrt{K}}\norm{\vect{x}}_1  \right).
\end{eqnarray}
When $\vect{x}$ is not exactly $K$-sparse we get that the effective
error in our results becomes $\tilde{\vect{e}} = \vect{e} +
\matr{D}(\vect{x} - \vect{x}_K)$. Thus, using the error bounds of
the algorithms with the inequality in (\ref{eq:near_approx_bound})
we get
\begin{eqnarray}
\norm{\vect{x} - \hat{\vect{x}}}_2 &\le& C
\norm{\matr{D}_{T_\vect{e}}^*\vect{\tilde{e}}}_2 \\
\nonumber
& \le &  C\norm{\matr{D}_{T_\vect{e}}^*\left(\vect{e} +
\matr{D}(\vect{x} - \vect{x}_K)\right)}_2 \\
\nonumber
& \le &  C\norm{\matr{D}_{T_\vect{e}}^*\vect{e}} +
C\norm{\matr{D}_{T_\vect{e}}^*\matr{D}(\vect{x} - \vect{x}_K)}_2 \\
\nonumber
& \le &  C\norm{\matr{D}_{T_\vect{e}}^*\vect{e}} +
C\sqrt{1+\delta_{K}}\norm{\matr{D}(\vect{x} - \vect{x}_K)}_2 \\
\nonumber & \le & C\bigg( \norm{\matr{D}_{T_\vect{e}}^*\vect{e}}_2
+ (1+\delta_K)\norm{\vect{x} - \vect{x}_K}_2   \\ \nonumber && +
\frac{1+\delta_K}{\sqrt{K}}\norm{\vect{x} - \vect{x}_K}_1 \bigg),
\end{eqnarray}
which proves (\ref{eq:approx_error_bound_consts}). Using the same
steps taken in Theorems \ref{thm:SP_oracle_thm},
\ref{thm:CoSaMP_oracle_thm}, and \ref{thm:IHT_oracle_thm}, lead us
to
\begin{IEEEeqnarray}{rCl} 
\label{eq:approx_error_bound_near_oracle_proof}  \norm{\vect{x} -
\hat{\vect{x}}}_2^2 &\le&  C^2\bigg( \sqrt{(2(1+a)\log{N})\cdot K}
\cdot \sigma  \\ \nonumber &&+ (1+\delta_{K})\norm{\vect{x} - \vect{x}_K}_2 +
\frac{1+\delta_{K}}{\sqrt{K}}\norm{\vect{x} - \vect{x}_K}_1
\bigg)^2 .
\end{IEEEeqnarray}
Since the RIP condition for all the algorithms satisfies $\delta_K
\le \sqrt{2} - 1$, plugging this into
(\ref{eq:approx_error_bound_near_oracle_proof}) gives
(\ref{eq:approx_error_bound_near_oracle}), and this concludes the
proof. \hfill $\Box$
\bigskip

Just as before, we should wonder how close is this bound to the one
obtained by an oracle that knows the support $T$ of the $K$ dominant
entries in $\vect{x}$. Following \cite{Blumensath09Iterative}, we
derive an expression for such an oracle. Using the fact that the
oracle is given by $\matr{D}_T^\dag \vect{y} = \matr{D}_T^\dag
(\matr{D}\vect{x}_T+\matr{D}(\vect{x}-\vect{x}_T)+\vect{e})$, its
MSE is bounded by
\begin{IEEEeqnarray}{rcl}
&&E\norm{\vect{x} - \hat{\vect{x}}_{oracle}}_2^2 = E\norm{\vect{x} -
\matr{D}_T^\dag \vect{y}}_2^2 \\
\nonumber  && =  E\norm{\vect{x} - \vect{x}_T -\matr{D}_T^\dag
\vect{e} - \matr{D}_T^\dag
\matr{D}(\vect{x} - \vect{x}_T)}_2^2 \\
\nonumber && \le  \left( \norm{\vect{x} - \vect{x}_T}_2 + \norm{
\matr{D}_T^\dag \matr{D}(\vect{x} - \vect{x}_T)}_2 +
E\norm{\matr{D}_T^\dag \vect{e} }_2 \right)^2,
\end{IEEEeqnarray}
where we have used the triangle inequality. Using the relation given
in (\ref{eq:oracle_perf2}) for the last term, and
properties of the RIP for the second, we obtain
\begin{IEEEeqnarray}{rcl} 
&&E\norm{\vect{x} - \hat{\vect{x}}_{oracle}}_2^2 \le  \\ \nonumber && ~~\left(
\norm{\vect{x} - \vect{x}_T}_2 + \frac{1}{\sqrt{1-\delta_{K}}}\norm{
\matr{D}(\vect{x} - \vect{x}_T)}_2 +
\frac{\sqrt{K}}{\sqrt{1-\delta_K}}\sigma \right)^2.
\end{IEEEeqnarray}
Finally, the middle-term can be further handled using
(\ref{eq:near_approx_bound}), and we arrive to
\begin{IEEEeqnarray}{rCl}
E\norm{\vect{x} - \hat{\vect{x}}_{oracle}}_2^2  &\le& \frac{1}{1-\delta_k}\bigg( (1
+\sqrt{1+\delta_{K}}) \norm{\vect{x} - \vect{x}_T}_2  \\ \nonumber &&+
\frac{\sqrt{1+\delta_{K}}}{\sqrt{K}} \norm{\vect{x} - \vect{x}_K}_1
+ \sqrt{K}\sigma \bigg)^2.
\end{IEEEeqnarray}
Thus we see again  that the error bound in the non-exact sparse
case, is up to a constant and the $\log{N}$ factor the same as the
one of the oracle estimator.


\section{Conclusion}
\label{sec:conc} In this paper we have presented near-oracle
performance guarantees for three greedy-like algorithms -- the
Subspace Pursuit, the CoSaMP, and the Iterative Hard-Thresholding.
The approach taken in our analysis is an RIP-based (as opposed to
mutual-coherence ones). Despite their resemblance to greedy
algorithms, such as the OMP and the thresholding, our study leads to
uniform guarantees for the three algorithms explored, i.e., the
near-oracle error bounds are dependent only on the dictionary
properties (RIP constant) and the sparsity level of the sought
solution. We have also presented a simple extension of our results
to the case where the representations are only approximately sparse.


\section*{Acknowledgment}

The authors would like to thank Zvika Ben-Haim for fruitful
discussions and relevant references to existing literature, which
helped in shaping this work.


\appendices

\section{Proof of Theorem \ref{thm:SP-1} -- inequality (\ref{eq:SP_x_diff_bound})}
\label{sec:SP_x_diff_bound_proof}

In the proof of (\ref{eq:SP_x_diff_bound}) we use two main inequalities:
\begin{eqnarray}
\label{eq:x_diff_tilde_prev_bound}
\norm{\vect{x}_{T-\tilde{T}^\ell}}_2 &\le& \frac{2\delta_{3K}}
{(1-\delta_{3K})^2}\norm{\vect{x}_{T-{T}^{\ell-1}}}_2  \\ \nonumber &&+
 \frac{2}{(1-\delta_{3K})^2}
\norm{\matr{D}_{T_\vect{e}}^*\vect{e}}_2,
\end{eqnarray}
and
\begin{eqnarray}
\label{eq:x_diff_curr_tilde_bound} \norm{\vect{x}_{T-T^\ell}}_2 &\le&
\frac{1+\delta_{3K}}{1-\delta_{3K}}\norm{\vect{x}_{T-\tilde{T}^\ell}}_2 \\ \nonumber &&
+ \frac{4}{1-\delta_{3K}}\norm{\matr{D}_{T_{\vect{e}}}^*\vect{e}}_2.
\end{eqnarray}
Their proofs are in Appendices
\ref{sec:x_diff_tilde_prev_bound_proof} and
\ref{sec:x_diff_curr_tilde_bound_proof} respectively. The inequality
(\ref{eq:SP_x_diff_bound}) is obtained by substituting
(\ref{eq:x_diff_tilde_prev_bound}) into
(\ref{eq:x_diff_curr_tilde_bound}) as shown below:
\begin{IEEEeqnarray}{rCl}
\norm{\vect{x}_{T-T^\ell}}_2 &\le& \frac{1+\delta_{3K}}
{1-\delta_{3K}}\norm{\vect{x}_{T-\tilde{T}^{\ell}}}_2+\frac{4}
{1-\delta_{3K}}\norm{\matr{D}_{T_{\vect{e}}}^*\vect{e}}_2 \\
\nonumber
&\le&  \frac{1+\delta_{3K}}{1-\delta_{3K}}\left[\frac{2\delta_{3K}}
{(1-\delta_{3K})^2}\norm{\vect{x}_{T-T^{\ell-1}}}_2\right. \\{} \nonumber  && \left.+ \frac{2}
{(1-\delta_{3K})^2}\norm{\matr{D}_{T_\vect{e}}^*\vect{e}}_2 \right]+
\frac{4}{1-\delta_{3K}}\norm{\matr{D}_{T_{\vect{e}}}^*\vect{e}}_2  \\
\nonumber
&\le& \frac{2\delta_{3K}(1+\delta_{3K})}{(1-\delta_{3K})^3}
\norm{\vect{x}_{T-T^{\ell-1}}}_2  \\ \nonumber  &&+\frac{2(1+\delta_{3K})}
{(1-\delta_{3K})^3}\norm{\matr{D}_{T_\vect{e}}^*\vect{e}}_2 +\frac{4}
{1-\delta_{3K}}\norm{\matr{D}_{T_{\vect{e}}}^*\vect{e}}_2
\\ \nonumber &\le&
\frac{2\delta_{3K}(1+\delta_{3K})}{(1-\delta_{3K})^3}
\norm{\vect{x}_{T-T^{\ell-1}}}_2 \\ \nonumber &&+\frac{6 - 6\delta_{3K}  + 4\delta_{3K}^2}
{(1-\delta_{3K})^3}
\norm{\matr{D}_{T_\vect{e}}^*\vect{e}}_2,
\end{IEEEeqnarray}
and this concludes this proof. \hfill $\Box$


\section{Proof of inequality (\ref{eq:x_diff_tilde_prev_bound})}
\label{sec:x_diff_tilde_prev_bound_proof}

\begin{lem}\label{lemma:A1}
The following inequality holds true for the SP algorithm:
\begin{eqnarray}
\nonumber \norm{\vect{x}_{T-\tilde{T}^\ell}}_2 &\le& \frac{2\delta_{3K}}
{(1-\delta_{3K})^2}\norm{\vect{x}_{T-{T}^{\ell-1}}}_2  \\ \nonumber &&+
 \frac{2}{(1-\delta_{3K})^2}
\norm{\matr{D}_{T_\vect{e}}^*\vect{e}}_2,
\end{eqnarray}
\end{lem}

{\em Proof:} We start by the residual-update step in the SP
algorithm, and exploit the relation $\vect{y} =
\matr{D}\vect{x}+\vect{e} =
\matr{D}_{T-T^{\ell-1}}\vect{x}_{T-T^{\ell-1}} + \matr{D}_{T\cap
T^{\ell-1}}\vect{x}_{T\cap T^{\ell-1}} +\vect{e}$. This leads to
\begin{IEEEeqnarray}{rCl}
\label{eq:resid_l1_composition_first}
\vect{y}_r^{\ell - 1} &=& \resid(\vect{y},\matr{D}_{T^{\ell-1}}) \\
\nonumber
&=& \resid(\matr{D}_{T-T^{\ell-1}}\vect{x}_{T-T^{\ell-1}},\matr{D}_{T^{\ell-1}}) \\
\nonumber
&& + \resid(\matr{D}_{T\cap T^{\ell-1}}
\vect{x}_{T\cap T^{\ell-1}},\matr{D}_{T^{\ell-1}})
+ \resid(\vect{e},\matr{D}_{T^{\ell-1}}).
\end{IEEEeqnarray}
Here we have used the linearity of the operator $\resid(\cdot,
\matr{D}_{T^{\ell-1}})$ with respect to its first entry. The second
term in the right-hand-side (rhs) is $0$ since $\matr{D}_{T\cap
T^{\ell-1}}\vect{x}_{T\cap T^{\ell-1}}\in
\spanned(\matr{D}_{T^{\ell-1}})$. For the first term in the rhs we
have
\begin{IEEEeqnarray}{rcl} 
\label{eq:resid_D_T_l_xr}
&&\resid(\matr{D}_{T-T^{\ell-1}}\vect{x}_{T-T^{\ell-1}},\matr{D}_{\ell-1}) \\ \nonumber
&& ~~=  \matr{D}_{T-T^{\ell-1}}\vect{x}_{T-T^{\ell-1}} -
\proj(\matr{D}_{T-T^{\ell-1}}\vect{x}_{T-T^{\ell-1}},\matr{D}_{T^{\ell-1}}) \\
\nonumber
&& ~~= \matr{D}_{T-T^{\ell-1}}\vect{x}_{T-T^{\ell-1}}  +
\matr{D}_{T^{\ell-1}}\vect{x}_{p,T^{\ell - 1}} \\
&& ~~ =   \left[ \matr{D}_{T-T^{\ell-1}},\matr{D}_{T^{\ell-1}}  \right]
\left[ \begin{array}{c}
\vect{x}_{T-T^{\ell-1}} \\
\nonumber \vect{x}_{p,T^{\ell - 1}} \end{array}\right]  \triangleq
\matr{D}_{T\cup T^{\ell-1}}\vect{x}_r^{\ell-1},
\end{IEEEeqnarray}
where we have defined
\begin{IEEEeqnarray}{rCl} 
\label{eq:xptdef}
\vect{x}_{p,T^{\ell - 1}} &=& -(\matr{D}^*_{T^{\ell -
1}}\matr{D}_{T^{\ell - 1}})^{-1}\matr{D}^*_{T^{\ell - 1}}
\matr{D}_{T-T^{\ell-1}}\vect{x}_{T-T^{\ell-1}}.
\end{IEEEeqnarray}
Combining (\ref{eq:resid_l1_composition_first}) and
(\ref{eq:resid_D_T_l_xr}) leads to
\begin{eqnarray}\label{eq:yrldef}
\vect{y}_r^{\ell - 1} = \matr{D}_{T\cup
T^{\ell-1}}\vect{x}_r^{\ell-1} +
\resid(\vect{e},\matr{D}_{T^{\ell-1}}).
\end{eqnarray}
By the definition of $T_\Delta$ in Algorithm \ref{alg:SP} we obtain
\begin{IEEEeqnarray}{rcl} 
\label{eq:x_diff_tilde_prev_bound_lower_bound_step1}
&& \norm{\matr{D}^*_{T_\Delta}\vect{y}_r^{\ell - 1}}_2  \ge
\norm{\matr{D}^*_T\vect{y}_r^{\ell - 1}}_2   \ge
\norm{\matr{D}^*_{T-T^{\ell-1}}\vect{y}_r^{\ell - 1}}_2\\
\nonumber &&
~~ \ge  \norm{\matr{D}^*_{T-T^{\ell-1}}
  \matr{D}_{T\cup T^{\ell - 1}}\vect{x}_r^{\ell - 1}}_2 \\ \nonumber && ~~~~ -
  \norm{\matr{D}^*_{T-T^{\ell-1}} \resid(\vect{e},\matr{D}_{T^{\ell-1}})}_2  \\
\nonumber &&
~~ \ge   \norm{\matr{D}^*_{T-T^{\ell-1}}
 \matr{D}_{T\cup T^{\ell - 1}}\vect{x}_r^{\ell - 1}}_2  -
 \norm{\matr{D}^*_{T-T^{\ell-1}} \vect{e}}_2
 \\ \nonumber && ~~~~- \norm{\matr{D}^*_{T-T^{\ell-1}} \proj(\vect{e},\matr{D}_{T^{\ell-1}})}_2.
\end{IEEEeqnarray}
We will bound $\norm{\matr{D}^*_{T-T^{\ell-1}}
\proj(\vect{e},\matr{D}_{T^{\ell-1}})}_2$ from above using RIP
properties from Section \ref{sec:notation},
\begin{eqnarray}
\label{eq:x_diff_tilde_prev_bound_lower_bound_step2}
&& \norm{\matr{D}^*_{T-T^{\ell-1}} \proj(\vect{e},\matr{D}_{T^{\ell-1}})}_2 \\ \nonumber
&& ~~= \norm{\matr{D}^*_{T-T^{\ell-1}} \matr{D}_{T^{\ell-1}}(\matr{D}_{T^{\ell-1}}^*
\matr{D}_{T^{\ell-1}})^{-1}\matr{D}_{T^{\ell-1}}^*\vect{e}}_2 \\
\nonumber
&&~~\le\frac{\delta_{3K}}{1-\delta_{3K}}\norm{\matr{D}_{T^{\ell-1}}^*\vect{e}}_2.
\end{eqnarray}
Combining (\ref{eq:x_diff_tilde_prev_bound_lower_bound_step1}) and
(\ref{eq:x_diff_tilde_prev_bound_lower_bound_step2}) leads to
\begin{IEEEeqnarray}{rCl} 
\label{eq:x_diff_tilde_prev_bound_lower_bound}
&& \norm{\matr{D}^*_{T_\Delta}\vect{y}_r^{\ell - 1}}_2
 \ge\norm{\matr{D}^*_{T-T^{\ell-1}}\matr{D}_{T\cup T^{\ell - 1}}
\vect{x}_r^{\ell - 1}}_2 \\ \nonumber &&~~~~~~~~~~~~~~ -  \norm{\matr{D}^*_{T} \vect{e}}_2 -
\frac{\delta_{3K}}{1-\delta_{3K}}\norm{\matr{D}_{T^{\ell-1}}^*\vect{e}}_2 \\
\nonumber && ~~~~~~~~ \ge\norm{\matr{D}^*_{T-T^{\ell-1}}
\matr{D}_{T\cup T^{\ell - 1}}\vect{x}_r^{\ell - 1}}_2  -
\frac{1}{1-\delta_{3K}}\norm{\matr{D}_{T_\vect{e}}^*\vect{e}}_2.
\end{IEEEeqnarray}
By the definition of $T_\Delta$ and $\vect{y}_r^{\ell-1}$ it holds
that $T_\Delta \cap T^{\ell-1} = \emptyset$ since
$\matr{D}_{T^{\ell-1}}^*\vect{y}_r^{\ell-1} = 0$. Using
(\ref{eq:yrldef}), the left-hand-side (lhs) of
(\ref{eq:x_diff_tilde_prev_bound_lower_bound}) is upper bounded by
\begin{IEEEeqnarray}{rcl} 
\label{eq:x_diff_tilde_prev_bound_upper_bound}
\norm{\matr{D}^*_{T_\Delta}\vect{y}_r^{\ell - 1}}_2 \\ \nonumber  && \le
\norm{\matr{D}^*_{T_\Delta}\matr{D}_{T\cup T^{\ell - 1}}\vect{x}_r^{\ell - 1}}_2 +
\norm{\matr{D}^*_{T_\Delta}\resid(\vect{e},\matr{D}_{T^{\ell-1}})}_2\\
\nonumber &&
 \le  \norm{\matr{D}^*_{T_\Delta}
\matr{D}_{T\cup T^{\ell - 1}}\vect{x}_r^{\ell - 1}}_2 +
\norm{\matr{D}^*_{T_\Delta}\vect{e}}_2
 \\ \nonumber && ~~~~
+ \norm{\matr{D}^*_{T_\Delta} \matr{D}_{T^{\ell-1}}
(\matr{D}_{T^{\ell-1}}^*\matr{D}_{T^{\ell-1}})^{-1}
\matr{D}_{T^{\ell-1}}^*\vect{e}}_2\\
\nonumber  && \le \norm{\matr{D}^*_{T_\Delta}\matr{D}_{T\cup T^{\ell
- 1}}\vect{x}_r^{\ell - 1}}_2 +
\norm{\matr{D}^*_{T_\Delta}\vect{e}}_2 \\ \nonumber && ~~~~+
\frac{\delta_{3K}}{1-\delta_{3K}}
\norm{\matr{D}_{T^{\ell-1}}^*\vect{e}}_2\\
\nonumber  && \le  \norm{\matr{D}^*_{T_\Delta}\matr{D}_{T\cup T^{\ell
- 1}}\vect{x}_r^{\ell - 1}}_2 +
\frac{1}{1-\delta_{3K}}\norm{\matr{D}^*_{T_\vect{e}}\vect{e}}_2.
\end{IEEEeqnarray}
Combining (\ref{eq:x_diff_tilde_prev_bound_lower_bound}) and
(\ref{eq:x_diff_tilde_prev_bound_upper_bound}) gives
\begin{eqnarray}
&&\norm{\matr{D}^*_{T_\Delta}\matr{D}_{T\cup T^{\ell -
1}}\vect{x}_r^{\ell - 1}}_2 +
\frac{2}{1-\delta_{3K}}\norm{\matr{D}_{T_\vect{e}}^*\vect{e}}_2 \\
\nonumber && ~~~~~~\ge
\norm{\matr{D}^*_{T-T^{\ell - 1}}\matr{D}_{T\cup T^{\ell -
1}}\vect{x}_r^{\ell - 1}}_2.
\end{eqnarray}
Removing the common rows in $\matr{D}^*_{T_\Delta}$ and
$\matr{D}^*_{T-T^{\ell - 1}}$ we get
\begin{IEEEeqnarray}{rCl} 
\label{eq:first_inequality_only_error_bounded}
&& \norm{\matr{D}^*_{T_\Delta - T}\matr{D}_{T\cup T^{\ell - 1}}
 \vect{x}_r^{\ell - 1}}_2  + \frac{2}{1-\delta_{3K}}
 \norm{\matr{D}_{T_\vect{e}}^*\vect{e}}_2 \\
\nonumber
&& ~~~~\ge
\norm{\matr{D}^*_{T-T^{\ell - 1}- T_{\Delta}}
\matr{D}_{T\cup T^{\ell - 1}}\vect{x}_r^{\ell - 1}}_2  \\
\nonumber
&& ~~~~=
\norm{\matr{D}^*_{T- \tilde{T}^{\ell}}
\matr{D}_{T\cup T^{\ell - 1}}\vect{x}_r^{\ell - 1}}_2  .
\end{IEEEeqnarray}
The last equality is true because
$T-T^{\ell - 1}- T_{\Delta} = T-T^{\ell - 1}-
(\tilde{T}^{\ell}-T^{\ell - 1}) = T - \tilde{T}^{\ell}$.

Now we turn to bound the lhs and rhs terms of
(\ref{eq:first_inequality_only_error_bounded}) from below and above,
respectively. For the lhs term we exploit the fact that the supports
$T_\Delta - T$ and $T \cup T^{\ell-1}$ are disjoint, leading to
\begin{IEEEeqnarray}{rCl} 
 \norm{\matr{D}^*_{T_\Delta - T}\matr{D}_{T\cup T^{\ell - 1}}
 \vect{x}_r^{\ell - 1}}_2  &\le&
\delta_{\abs{T_\Delta\cup T^{\ell-1}\cup T}}\norm{\vect{x}^{\ell-1}_r}_2
\\ \nonumber &\le& \delta_{3K}\norm{\vect{x}^{\ell-1}_r}_2
\end{IEEEeqnarray}
For the rhs term in (\ref{eq:first_inequality_only_error_bounded}),
we obtain
\begin{IEEEeqnarray}{rcl} 
&& \norm{\matr{D}^*_{T - \tilde{T}^{\ell}}
 \matr{D}_{T\cup T^{\ell - 1}}\vect{x}_r^{\ell - 1}}_2 \\ \nonumber &&
~~\ge  \norm{\matr{D}^*_{T - \tilde{T}^{\ell}}\matr{D}_{T - \tilde{T}^{\ell}}
 (\vect{x}_r^{\ell - 1})_{T - \tilde{T}^{\ell}}}_2 \\
\nonumber && ~~~~-
 \norm{\matr{D}^*_{T - \tilde{T}^{\ell}}\matr{D}_{(T\cup T^{\ell - 1}) -
 (T - \tilde{T}^{\ell})}(\vect{x}_r^{\ell - 1})_{(T\cup T^{\ell - 1})
 - (T - \tilde{T}^{\ell})}}_2 \\
\nonumber
&& ~~\ge (1-\delta_{K})\norm{(\vect{x}_r^{\ell - 1})_{T - \tilde{T}^{\ell}}}_2
- \delta_{3K}\norm{\vect{x}_r^{\ell - 1}}_2 \\
\nonumber
&& ~~\ge (1-\delta_{3K})\norm{(\vect{x}_r^{\ell - 1})_{T - \tilde{T}^{\ell}}}_2
- \delta_{3K}\norm{\vect{x}_r^{\ell - 1}}_2
\end{IEEEeqnarray}
Substitution of the two bounds derived above into
(\ref{eq:first_inequality_only_error_bounded}) gives
\begin{eqnarray}
\label{eq:first_inequality_only_error_bounded_x_r}
&& 2\delta_{3K}\norm{\vect{x}^{\ell-1}_r}_2  +
\frac{2}{1-\delta_{3K}}\norm{\matr{D}_{T_\vect{e}}^*\vect{e}}_2 \\ \nonumber && ~~~~~~\ge
(1-\delta_{3K})\norm{(\vect{x}_r^{\ell - 1})_{T - \tilde{T}^{\ell}}}_2.
\end{eqnarray}
The above inequality uses $\vect{x}^{\ell-1}_r$, which was defined
in (\ref{eq:resid_D_T_l_xr}), and this definition relies on
yet another one definition for the vector $\vect{x}_{p,T^{\ell -
1}}$ in (\ref{eq:xptdef}). We proceed by bounding
$\norm{\vect{x}_{p,T^{\ell - 1}}}_2$ from above,
\begin{IEEEeqnarray}{rCl} 
\label{eq:x_p_T_upper_bound}
&& \norm{\vect{x}_{p,T^{\ell - 1}}}_2 \\ \nonumber
&& ~~= \norm{-(\matr{D}^*_{T^{\ell - 1}}
\matr{D}_{T^{\ell - 1}})^{-1}\matr{D}^*_{T^{\ell - 1}}
\matr{D}_{T-T^{\ell-1}}\vect{x}_{T-T^{\ell-1}}}_2 \\
\nonumber
&& ~~\le  \frac{1}{1-\delta_K}\norm{-\matr{D}^*_{T^{\ell - 1}}
\matr{D}_{T-T^{\ell-1}}\vect{x}_{T-T^{\ell-1}}}_2\\
\nonumber &&~~\le
\frac{\delta_{2K}}{1-\delta_K}\norm{\vect{x}_{T-T^{\ell-1}}}_2 \le
\frac{\delta_{3K}}{1-\delta_{3K}}\norm{\vect{x}_{T-T^{\ell-1}}}_2,
\end{IEEEeqnarray}
and get
\begin{eqnarray}
\label{eq:x_r_l_norm_upper_bound}
\norm{\vect{x}_r^{\ell-1}}_2 &\le& \norm{\vect{x}_{T-T^{\ell-1}}}_2 +
\norm{\vect{x}_{p,T^{\ell-1}}}_2  \\ \nonumber &\le& \left( 1+ \frac{\delta_{3K}}
{1-\delta_{3K}}\right) \norm{\vect{x}_{T-T^{\ell-1}}}_2 \\
\nonumber
&\le& \frac{1}{1-\delta_{3K}} \norm{\vect{x}_{T-T^{\ell-1}}}_2.
\end{eqnarray}
In addition, since $(\vect{x}_r^{\ell - 1})_{T - T^{\ell-1}} =
\vect{x}_{T - T^{\ell-1}}$ then $(\vect{x}_r^{\ell - 1})_{T -
\tilde{T}^{\ell}} =\vect{x}_{T - \tilde{T}^{\ell}}$. Using this fact
and (\ref{eq:x_r_l_norm_upper_bound})  with
(\ref{eq:first_inequality_only_error_bounded_x_r}) leads to
\begin{eqnarray}
&& \norm{\vect{x}_{T-\tilde{T}^\ell}}_2 \\ \nonumber
&& ~~\le \frac{2\delta_{3K}}
{(1-\delta_{3K})^2}\norm{\vect{x}_{T-{T}^{\ell-1}}}_2  +
\frac{2}{(1-\delta_{3K})^2}\norm{\matr{D}_{T_\vect{e}}^*\vect{e}}_2,
\end{eqnarray}
which proves the inequality in (\ref{eq:x_diff_tilde_prev_bound}).
\hfill $\Box$


\section{Proof of inequality (\ref{eq:x_diff_curr_tilde_bound})}
\label{sec:x_diff_curr_tilde_bound_proof}

\begin{lem}\label{lemma:A2}
The following inequality holds true for the SP algorithm:
\begin{eqnarray}
\nonumber \norm{\vect{x}_{T-T^\ell}}_2 &\le &
\frac{1+\delta_{3K}}{1-\delta_{3K}}\norm{\vect{x}_{T-\tilde{T}^\ell}}_2
+ \frac{4}{1-\delta_{3K}}\norm{\matr{D}_{T_{\vect{e}}}^*\vect{e}}_2.
\end{eqnarray}
\end{lem}

{\em Proof:} We will define the smear vector $\epsilonbf =
\vect{x}_p - \vect{x}_{\tilde{T}^\ell}$, where $\vect{x}_p$ is the
outcome of the representation computation over ${\tilde T}^{\ell}$,
given by
\begin{eqnarray}
\label{eq:SP_x_p_explicit}
\vect{x_p} = \matr{D}_{\tilde{T}^\ell}^\dag\vect{y} =
\matr{D}_{\tilde{T}^\ell}^\dag(\matr{D}_T\vect{x}_T + \vect{e}),
\end{eqnarray}
as defined in Algorithm~\ref{alg:SP}. Expanding the first term in the last equality gives:
\begin{IEEEeqnarray}{rCl} 
\label{eq:SP_x_p_explicit_first_element}
&& \matr{D}_{\tilde{T}^\ell}^\dag\matr{D}_T\vect{x}_T =
 \matr{D}_{\tilde{T}^\ell}^\dag \matr{D}_{T\cap \tilde{T}^\ell}\vect{x}_{T\cap \tilde{T}^\ell} +
 \matr{D}_{\tilde{T}^\ell}^\dag \matr{D}_{T- \tilde{T}^\ell}\vect{x}_{T- \tilde{T}^\ell} \\
\nonumber
&& ~ =  \matr{D}_{\tilde{T}^\ell}^\dag \left[ \matr{D}_{T\cap \tilde{T}^\ell},
\matr{D}_{\tilde{T}^\ell - T} \right]  \left[ \begin{array}{c}
\vect{x}_{T\cap \tilde{T}^\ell} \\
 \vect{0} \end{array}\right] +  \matr{D}_{\tilde{T}^\ell}^\dag
\matr{D}_{T- \tilde{T}^\ell}\vect{x}_{T- \tilde{T}^\ell} \\
\nonumber
&& ~=  \matr{D}_{\tilde{T}^\ell}^\dag \matr{D}_{\tilde{T}^\ell}
\vect{x}_{\tilde{T}^\ell}   +  \matr{D}_{\tilde{T}^\ell}^\dag
\matr{D}_{T- \tilde{T}^\ell}\vect{x}_{T- \tilde{T}^\ell} \\
\nonumber
&& ~= \vect{x}_{\tilde{T}^\ell}   +  \matr{D}_{\tilde{T}^\ell}^\dag
\matr{D}_{T- \tilde{T}^\ell}\vect{x}_{T- \tilde{T}^\ell}.
\end{IEEEeqnarray}
The equalities hold based on the definition of
$\matr{D}_{\tilde{T}^\ell}^\dag$ and on the fact that $\vect{x}$ is
$0$ outside of $T$. Using (\ref{eq:SP_x_p_explicit_first_element})
we bound the smear energy from  above, obtaining
\begin{eqnarray}
\label{eq:smear_upper_bound} \norm{\epsilonbf}_2 &\le&
\norm{\matr{D}_{\tilde{T}^\ell}^\dag\matr{D}_T\vect{x}_T
}_2 + \norm{\matr{D}_{\tilde{T}^\ell}^\dag\vect{e}}_2 \\
\nonumber
& = &
\norm{(\matr{D}_{\tilde{T}^\ell}^*\matr{D}_{\tilde{T}^\ell})^{-1}
\matr{D}_{\tilde{T}^\ell}^* \matr{D}_{T- \tilde{T}^\ell}\vect{x}_{T-
\tilde{T}^\ell}}_2 \\ \nonumber && + \norm{(\matr{D}_{\tilde{T}^\ell}^*
\matr{D}_{\tilde{T}^\ell})^{-1}\matr{D}_{\tilde{T}^\ell}^*\vect{e}}_2  \\
\nonumber
& \le &
\frac{\delta_{3K}}{1-\delta_{3K}}\norm{\vect{x}_{T- \tilde{T}^\ell}}_2 +
\frac{1}{1-\delta_{3K}}\norm{\matr{D}_{\tilde{T}^\ell}^*\vect{e}}_2.
\end{eqnarray}

We now turn to bound $\norm{\epsilonbf}_2$ from below. We denote the
support of the $K$ smallest coefficients in $\vect{x}_p$ by $\Delta
T \triangleq \tilde{T}^\ell - T^\ell$. Thus, for any set $T' \subset
\tilde{T}^\ell$ of cardinality $K$, it holds that
$\norm{(\vect{x}_p)_{\Delta T}}_2 \le \norm{(\vect{x}_p)_{T'}}_2$.
In particular, we shall choose $T'$ such that $T' \cap T =
\emptyset$, which necessarily exists because $\tilde{T}^\ell$ is of
cardinality $2K$ and therefore there must be at $K$ entries in this
support that are outside $T$. Thus, using the relation $\epsilonbf =
\vect{x}_p - \vect{x}_{\tilde{T}^\ell}$ we get
\begin{eqnarray}\label{eq:xpdeltat}
\norm{(\vect{x}_p)_{\Delta T}}_2 &\le&  \norm{(\vect{x}_p)_{T'}}_2 =
\norm{\left(\vect{x}_{\tilde{T}^\ell} \right)_{T'}+
\epsilonbf_{T'}}_2 \\ \nonumber &=& \norm{ \epsilonbf_{T'}}_2 \le \norm{
\epsilonbf}_2.
\end{eqnarray}
Because $\vect{x}$ is supported on $T$ we have that
$\norm{\vect{x}_{\Delta T}}_2 = \norm{\vect{x}_{\Delta T \cap
T}}_2$. An upper bound for this vector is reached by
\begin{eqnarray}
\label{eq:smear_lower_bound} \norm{\vect{x}_{\Delta T \cap T}}_2 &=&
\norm{(\vect{x}_p)_{\Delta T \cap T} -\epsilonbf_{\Delta T \cap
T}}_2 \\ \nonumber  &\le& \norm{(\vect{x}_p)_{\Delta T \cap T}}_2 +
\norm{\epsilonbf_{\Delta T \cap T}}_2 \\
\nonumber & \le & \norm{(\vect{x}_p)_{\Delta T}}_2 +
\norm{\epsilonbf}_2 \le 2 \norm{\epsilonbf}_2,
\end{eqnarray}
where the last step uses (\ref{eq:xpdeltat}). The vector
$\vect{x}_{T-T^\ell}$ can be decomposed as $\vect{x}_{T-T^\ell} =
\left[ \vect{x}_{T\cap \Delta T}^*,
\vect{x}^*_{T-\tilde{T}^\ell}\right]^*$. Using
(\ref{eq:smear_upper_bound}) and (\ref{eq:smear_lower_bound}) we get
\begin{IEEEeqnarray}{c} 
\nonumber
\norm{\vect{x}_{T-T^\ell}}_2 \le  \norm{\vect{x}_{T\cap \Delta T}}_2 +
\norm{\vect{x}_{T-\tilde{T}^\ell}}_2 \le 2\norm{\epsilonbf}_2 + \norm{\vect{x}_{T-\tilde{T}^\ell}}_2 \\
\nonumber
~~ \le  \left(1 + \frac{2\delta_{3K}}{1-\delta_{3K}}\right)
\norm{\vect{x}_{T-\tilde{T}^\ell}}_2 + \frac{2}
{1-\delta_{3K}}\norm{\matr{D}_{\tilde{T}^\ell}^*\vect{e}}_2\\
\nonumber  ~~=
\frac{1+\delta_{3K}}{1-\delta_{3K}}\norm{\vect{x}_{T-\tilde{T}^\ell}}_2
+ \frac{4}{1-\delta_{3K}}\norm{\matr{D}_{T_{\vect{e}}}^*\vect{e}}_2,
\end{IEEEeqnarray}
where the last step uses the property $\norm{\matr{D}_{{\tilde
T}^{\ell}}^*\vect{e}}_2 \le
2\norm{\matr{D}_{T_{\vect{e}}}^*\vect{e}}_2$ taken from Section
\ref{sec:notation}, and this concludes the proof. \hfill $\Box$


\section{Proof of inequality (\ref{eq:CoSaMP_x_diff_bound})}
\label{sec:CoSaMP_x_diff_bound_proof}

\begin{lem}\label{lemma:D1}
The following inequality holds true for the CoSaMP algorithm:
\begin{eqnarray}
\nonumber \norm{\vect{x}-\hat{\vect{x}}_{CoSaMP}^\ell}_2  \le
\frac{4\delta_{4K}}{(1-\delta_{4K})^2}\norm{\vect{x}-\hat{\vect{x}}_{CoSaMP}^{\ell-1}}_2
+ \frac{14-6\delta_{4K} }{(1-\delta_{4K})^2}
\norm{\matr{D}_{T_{\vect{e}}}^*\vect{e}}_2.
\end{eqnarray}
\end{lem}

{\em Proof:} We denote $\hat{\vect{x}}_{CoSaMP}^\ell$ as the
solution of CoSaMP in the $\ell$-th iteration:
$\hat{\vect{x}}_{CoSaMP,(T^\ell)^C}^{\ell} = 0$ and
$\hat{\vect{x}}_{CoSaMP,T^\ell}^{\ell} = (\vect{x}_p)_{T^\ell}$. We
further define $\vect{r}^\ell \triangleq
\vect{x}-\hat{\vect{x}}_{CoSaMP}^\ell$ and use the definition of
$T_{\vect{e},p}$ (Section \ref{sec:notation}). Our proof is based on
the proof of Theorem 4.1 and the Lemmas used with it
\cite{Needell09CoSaMP}.

Since we choose $T_\Delta$ to contain the biggest $2K$ elements in
$\matr{D}^*\vect{y}_r^{\ell}$ and $\abs{T^{\ell - 1} \cup T}\le 2K$
it holds true that $ \norm{(\matr{D}^*\vect{y}_r^{\ell})_{T^{\ell }
\cup T}}_2 \le \norm{(\matr{D}^*\vect{y}_r^{\ell})_{T_\Delta}}_2 $.
Removing the common elements from both sides we get
\begin{equation}
\label{eq:CoSaMP_resid_basic_inequality}
\norm{(\matr{D}^*\vect{y}_r^{\ell})_{(T^{\ell } \cup T) - T_\Delta}}_2
\le \norm{(\matr{D}^*\vect{y}_r^{\ell})_{T_\Delta - (T^{\ell} \cup T)}}_2.
\end{equation}
We proceed by bounding the rhs and lhs of
(\ref{eq:CoSaMP_resid_basic_inequality}), from above and from  below
respectively, using the triangle inequality. We use Propositions
\ref{prop1} and \ref{prop2}, the definition of $T_{\vect{e},2}$, and
the fact that $\norm{\vect{r}^{\ell}}_2 = \norm{\vect{r}_{T^{\ell }
\cup T}^{\ell}}_2$ (this holds true since the support of
$\vect{r}^{\ell}$ is over $T \cup T^{\ell}$). For the rhs we obtain
\begin{eqnarray}
\label{eq:CoSaMP_resid_basic_inequality_rhs}
\norm{(\matr{D}^*\vect{y}_r^{\ell})_{T_\Delta - (T^{\ell} \cup T)}}_2 & = &
\norm{\matr{D}^*_{T_\Delta - (T^{\ell} \cup T)}(\matr{D}\vect{r}^{\ell}+\vect{e})}_2 \\
\nonumber
& \le &  \norm{\matr{D}^*_{T_\Delta - (T^{\ell} \cup T)}
\matr{D}_{T^{\ell} \cup T}\vect{r}_{T^{\ell} \cup T}^{\ell}}_2 +
\norm{\matr{D}^*_{T_\Delta - (T^{\ell} \cup T)}\vect{e}}_2 \\
\nonumber & \le &  \delta_{4K}\norm{\vect{r}^{\ell}}_2 +
\norm{\matr{D}^*_{T_{\vect{e},2}}\vect{e}}_2.
\end{eqnarray}
and for the lhs:
\begin{eqnarray}
\label{eq:CoSaMP_resid_basic_inequality_lhs}
\norm{(\matr{D}^*\vect{y}_r^{\ell})_{(T^{\ell} \cup T) - T_\Delta}}_2
&=&  \norm{\matr{D}^*_{(T^{\ell} \cup T) - T_\Delta}(\matr{D}\vect{r}^{\ell}+\vect{e})}_2 \\
\nonumber
&\ge & \norm{\matr{D}^*_{(T^{\ell} \cup T) -
T_\Delta}\matr{D}_{(T^{\ell} \cup T) - T_\Delta}
\vect{r}^{\ell}_{(T^{\ell} \cup T) - T_\Delta}}_2 \\
\nonumber &&- \norm{\matr{D}^*_{(T^{\ell} \cup T) - T_\Delta}
\matr{D}_{T_{\Delta}}\vect{r}^{\ell}_{T_{\Delta}}}_2 -
\norm{\matr{D}^*_{(T^{\ell} \cup T) - T_\Delta}\vect{e}}_2\\
\nonumber &\ge & (1-\delta_{2K})\norm{\vect{r}^{\ell}_{(T^{\ell}
\cup T) - T_\Delta}}_2 -
\delta_{4K}\norm{\vect{r}^{\ell}_{T_{\Delta}}}_2 -
\norm{\matr{D}^*_{T_{\vect{e},2}}\vect{e}}_2.
\end{eqnarray}
Because $\vect{r}^{\ell}$ is supported over $T \cup T^{\ell}$, it
holds true that $\norm{\vect{r}^\ell_{(T \cup T^{\ell}) -
T_\Delta}}_2 = \norm{\vect{r}^\ell_{T_\Delta^C}}_2$. Combining
(\ref{eq:CoSaMP_resid_basic_inequality_lhs}) and
(\ref{eq:CoSaMP_resid_basic_inequality_rhs}) with
(\ref{eq:CoSaMP_resid_basic_inequality}),  gives
\begin{eqnarray}
\label{eq:CoSaMP_identification}
\norm{\vect{r}^{\ell}_{T_\Delta^C}}_2 &\le&
\frac{2\delta_{4K}\norm{\vect{r}^{\ell}}_2+
2\norm{\matr{D}_{T_{\vect{e},2}}^*\vect{e}}_2}{1-\delta_{2K}} \\
&\le& \frac{2\delta_{4K}\norm{\vect{r}^{\ell}}_2+
4\norm{\matr{D}_{T_{\vect{e}}}^*\vect{e}}_2}{1-\delta_{4K}}.\nonumber
\end{eqnarray}

For brevity of notations, we denote hereafter ${\tilde T}^{\ell}$ as
${\tilde T}$. Using $\vect{y} = \matr{D}\vect{x} + \vect{e} =
\matr{D}_{\tilde{T}}\vect{x}_{\tilde{T}} +
\matr{D}_{\tilde{T}^C}\vect{x}_{\tilde{T}^C}  + \vect{e}$, we
observe that
\begin{eqnarray}
\norm{\vect{x}_{\tilde{T}} - (\vect{x}_p)_{\tilde{T}}}_2 &=&
\norm{\vect{x}_{\tilde{T}} - \matr{D}_{\tilde{T}}^\dag
\left(\matr{D}_{\tilde{T}}\vect{x}_{\tilde{T}} + \matr{D}_{\tilde{T}^C}\vect{x}_{\tilde{T}^C}  + \vect{e}\right)}_2 \\
\nonumber
&=& \norm{\matr{D}_{\tilde{T}}^\dag \left( \matr{D}_{\tilde{T}^C}\vect{x}_{\tilde{T}^C}  + \vect{e}\right)}_2 \\
\nonumber
& \le &  \norm{(\matr{D}_{\tilde{T}}^*\matr{D}_{\tilde{T}})^{-1} \matr{D}_{\tilde{T}}^* \matr{D}_{\tilde{T}^C}\vect{x}_{\tilde{T}^C}}_2 + \norm{(\matr{D}_{\tilde{T}}^*\matr{D}_{\tilde{T}})^{-1} \matr{D}_{\tilde{T}}^*  \vect{e}}_2 \\
\nonumber
&\le&  \frac{1}{1-\delta_{3K}}\norm{\matr{D}_{\tilde{T}}^* \matr{D}_{\tilde{T}^C}\vect{x}_{\tilde{T}^C}}_2 + \frac{1}{1-\delta_{3K}}\norm{\matr{D}_{T_{\vect{e},3}}^*\vect{e}}_2 \\
&\le&
\frac{\delta_{4K}}{1-\delta_{4K}}\norm{\vect{x}_{\tilde{T}^C}}_2 +
\frac{3}{1-\delta_{4K}}\norm{\matr{D}_{T_{\vect{e}}}^*\vect{e}}_2,
\nonumber
\end{eqnarray}
where the last inequality holds true because of  Proposition
\ref{prop4} and that $|\tilde{T}| = 3K$. Using the triangle
inequality and the fact that $\vect{x}_p$ is supported on ${\tilde
T}$, we obtain
\begin{eqnarray}
\norm{\vect{x}-\vect{x}_p}_2 \le \norm{\vect{x}_{\tilde{T}^C}}_2 +
\norm{\vect{x}_{\tilde{T}} - (\vect{x}_p)_{\tilde{T}}}_2,
\end{eqnarray}
which leads to
\begin{eqnarray}
\label{eq:CoSaMP_estimation}
\norm{\vect{x} - \vect{x}_p}_2 &\le&  \left( 1+ \frac{\delta_{4K}}{1-\delta_{4K}} \right) \norm{\vect{x}_{\tilde{T}^C}}_2  + \frac{3}{1-\delta_{4K}}\norm{\matr{D}_{T_{\vect{e}}}^*\vect{e}}_2 \\
  &=&  \frac{1}{1-\delta_{4K}}\norm{\vect{x}_{\tilde{T}^C}}_2  +
  \frac{3}{1-\delta_{4K}}\norm{\matr{D}_{T_{\vect{e}}}^*\vect{e}}_2.
  \nonumber
\end{eqnarray}
Having the above results we can obtain
(\ref{eq:CoSaMP_x_diff_bound}) by
\begin{eqnarray}
\label{eq:CoSaMP_bound_first_deriv}
\norm{\vect{x}-\hat{\vect{x}}_{CoSaMP}^\ell}_2 &\le&  2\norm{\vect{x}-\vect{x}_p}_2 \\
\nonumber
&\le&  2\left( \frac{1}{1-\delta_{4K}}\norm{\vect{x}_{\tilde{T}^C}}_2  + \frac{3}{1-\delta_{4K}}\norm{\matr{D}_{T_{\vect{e}}}^*\vect{e}}_2 \right) \\
\nonumber
&\le&  \frac{2}{1-\delta_{4K}} \norm{\vect{r}^{\ell-1}_{T_\Delta^C}}_2  + \frac{6}{1-\delta_{4K}}\norm{\matr{D}_{T_{\vect{e}}}^*\vect{e}}_2  \\
\nonumber
&\le&\frac{2}{1-\delta_{4K}} \left( \frac{2\delta_{4K}}{1-\delta_{4K}}\norm{\vect{r}^{\ell - 1}}_2+ \frac{4}{1-\delta_{4K}}\norm{\matr{D}_{T_{\vect{e}}}^*\vect{e}}_2 \right)  + \frac{6}{1-\delta_{4K}}\norm{\matr{D}_{T_{\vect{e}}}^*\vect{e}}_2 \\
\nonumber & = &
\frac{4\delta_{4K}}{(1-\delta_{4K})^2}\norm{\vect{r}^{\ell - 1}}_2
+ \frac{14-6\delta_{4K} }{(1-\delta_{4K})^2}
\norm{\matr{D}_{T_{\vect{e}}}^*\vect{e}}_2,
\end{eqnarray}
where the inequalities are based on Lemma 4.5 from
\cite{Needell09CoSaMP}, (\ref{eq:CoSaMP_estimation}), Lemma 4.3 from
\cite{Needell09CoSaMP} and (\ref{eq:CoSaMP_identification})
respectively. \hfill $\Box$


\linespread{1.3} \small{
\bibliographystyle{IEEEtran}
\bibliography{greedy_RIP_denoising}}

\end{document}